\PassOptionsToPackage{table}{xcolor}
\PassOptionsToPackage{prologue,dvipsnames}{xcolor}

\documentclass[manuscript]{acmart}
\usepackage{comment}
\usepackage{graphicx}
\usepackage{algorithm}
\usepackage{algpseudocode}
\usepackage{pifont}
\usepackage{amsmath}
\usepackage[font=bf]{caption}
\usepackage{makecell}
\usepackage[framemethod=TikZ]{mdframed}
\mdfsetup{skipabove=2pt}
\usepackage{fontawesome}
\usepackage{multirow}

\AtBeginDocument{%
  \providecommand\BibTeX{{%
    \normalfont B\kern-0.5em{\scshape i\kern-0.25em b}\kern-0.8em\TeX}}}

\begin{document}

\title{Leveraging Reviewer Experience in Code Review Comment Generation}

\author{Hong Yi Lin}
\email{holin2@unimelb.edu.au}
\orcid{0009-0004-5368-8897}
\affiliation{%
  \institution{The University of Melbourne}
  \city{Melbourne}
  \state{Victoria}
  \country{Australia}
}

\author{Patanamon Thongtanunam}
\email{patanamon.t@unimelb.edu.au}
\orcid{0000-0001-6328-8839}
\affiliation{%
  \institution{The University of Melbourne}
  \city{Melbourne}
  \state{Victoria}
  \country{Australia}
}

\author{Christoph Treude}
\email{ctreude@smu.edu.sg}
\orcid{0000-0002-6919-2149}
\affiliation{%
  \institution{Singapore Management University}
  \city{Singapore}
  \country{Singapore}
}

\author{Michael W. Godfrey}
\email{migod@uwaterloo.ca}
\orcid{0000-0001-5500-025X}
\affiliation{%
  \institution{University of Waterloo}
  \city{Waterloo}
  \state{Ontario}
  \country{Canada}
}

\author{Chunhua Liu}
\email{chunhua@student.unimelb.edu.au}
\orcid{0009-0009-6778-0172}
\affiliation{%
  \institution{The University of Melbourne}
  \city{Melbourne}
  \state{Victoria}
  \country{Australia}
}

\author{Wachiraphan Charoenwet}
\email{wcharoenwet@student.unimelb.edu.au}
\orcid{0000-0002-9814-3514}
\affiliation{%
  \institution{The University of Melbourne}
  \city{Melbourne}
  \state{Victoria}
  \country{Australia}
}

\renewcommand{\shortauthors}{Lin et al.}

\begin{abstract}

\end{abstract}

\ccsdesc[500]{Software and its engineering~Software creation and management}
\ccsdesc[500]{Computing methodologies~Machine translation}
\ccsdesc[500]{Computing methodologies~Natural language generation}
\keywords{Code Review, Review Comments, Neural Machine Translation, Natural Language Generation}

\begin{abstract}
Modern code review is a ubiquitous software quality assurance process aimed at identifying and resolving potential issues (e.g., functional, evolvability) within newly written code.
Despite its effectiveness, the process demands large amounts of effort from the human reviewers involved.
To help alleviate this workload, researchers have trained various deep learning based language models to imitate human reviewers in providing natural language code reviews for submitted code.
Formally, this automation task is known as code review comment generation.
Prior work has demonstrated improvements in code review comment generation by leveraging machine learning techniques and neural models, such as transfer learning and the transformer architecture.
However, the quality of the model generated reviews remain sub-optimal due to the quality of the open-source code review data used in model training.
This is in part due to the data obtained from open-source projects where code reviews are conducted in a public forum, and reviewers possess varying levels of software development experience, potentially affecting the quality of their feedback.
To accommodate for this variation, we propose a suite of experience-aware training methods that utilise the reviewers' past authoring and reviewing experiences as signals for review quality.
Specifically, we propose experience-aware loss functions (ELF), which use the reviewers' authoring and reviewing ownership of a project as weights in the model's loss function. 
Through this method, experienced reviewers' code reviews yield larger influence over the model's behaviour.
Compared to the SOTA model, ELF was able to generate higher quality reviews in terms of accuracy (e.g., +29\% applicable comments), informativeness (e.g., +56\% suggestions), and issue types discussed (e.g., +129\% functional issues identified).
The key contribution of this work is the demonstration of how traditional software engineering concepts such as reviewer experience can be integrated into the design of AI-based automated code review models.
\end{abstract}

\maketitle

\section{Introduction}
As a spiritual successor to the Fagan inspection, modern code review is a lightweight human-oriented software quality assurance process that can be found in most of today's collaborative software development environments~\cite{convergent2013,didact,vijayvergiya2024ai}.
The process requires developers who are not the code author to inspect a code change for a wide variety of problems, ranging from functional issues (e.g., logical flaws and resource misuse) to evolvability issues (e.g., poor variable naming and low readability)~\cite{mantyala}.
Whilst the review process helps improve the quality and maintainability of the software, practitioners often find the process both time-consuming~\cite{baum} and mentally taxing~\cite{cognitive2022}.
When code reviews are not rigorously conducted, software files are still found to be defective~\cite{7180077}, emphasising a need for automation.

To help alleviate this workload, researchers have attempted to automate the code review process via three sequential tasks: \textit{code change quality estimation}, \textit{code review comment generation}, and \textit{code refinement}.
Respectively, these tasks replicate the developer's actions in deciding if a code change needs to be revised, describing what needs to be revised in a natural language review, and finally revising the code according to that review.
In this work, we focus on the \textit{code review comment generation} task which has been shown to be the most challenging component of the three tasks~\cite{codereviewer}.
This involves training deep language models~\cite{vaswani2017attention,raffel2020exploring} to provide natural language reviews that identify a wide variety of issues within submitted code changes~\cite{tufano2022using,auger,codereviewer}.
The state-of-the-art model, CodeReviewer~\cite{codereviewer}, is a T5-based transformer~\cite{raffel2020exploring} that has been trained on a large-scale GitHub code review corpus.
Although considerable effort has been invested into curating a diverse set of code reviews, little attention has been placed on exploring the variation in quality across the reviews themselves i.e., all past studies including CodeReviewer have weighted the data uniformly during model training, allowing equal influence over model behaviour, despite potential differences in content quality.

\begin{figure}
    \centering
    \includegraphics[width=1\textwidth]{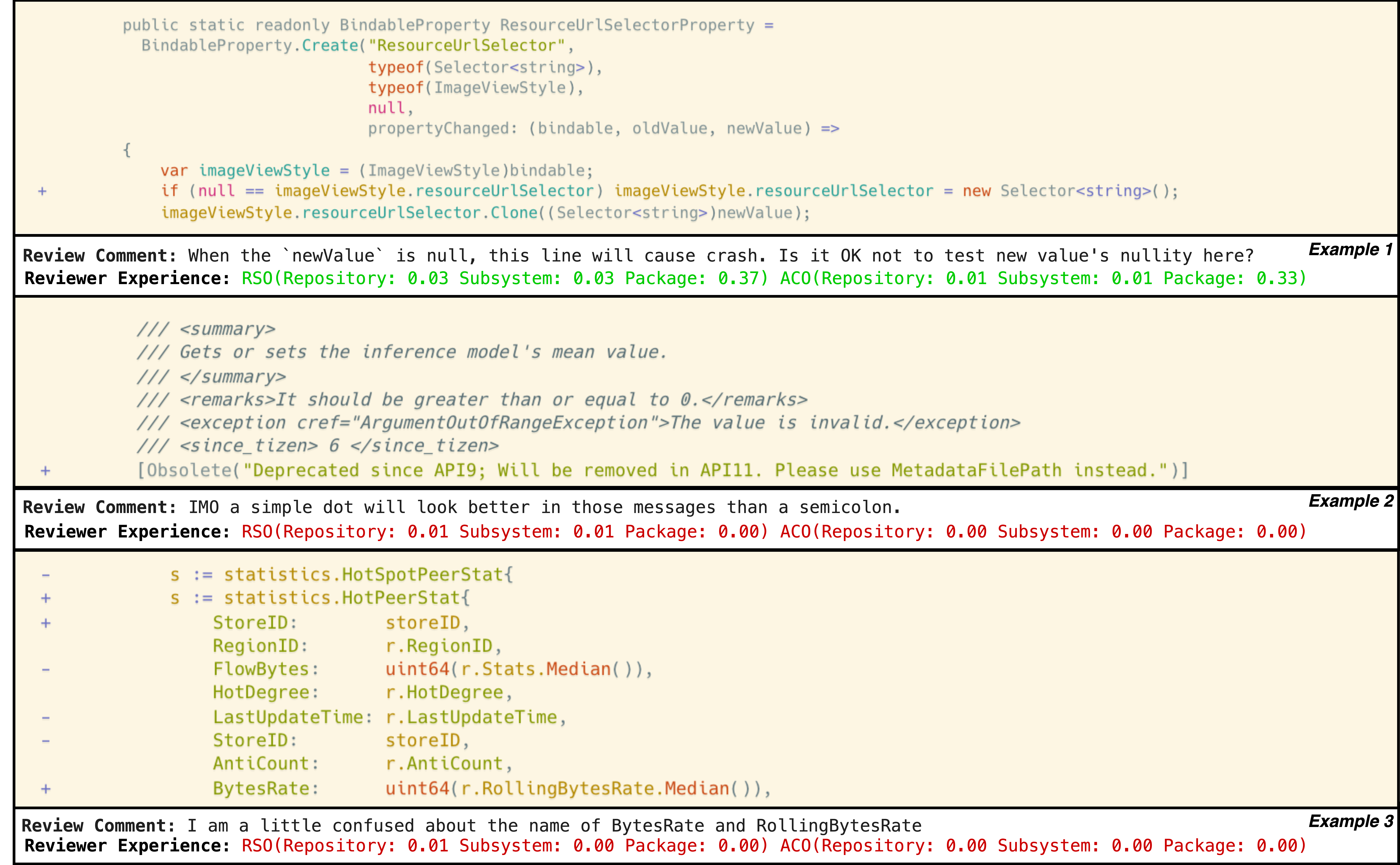}
    \caption{Example of Variation in Code Review Content Provided by Reviewers with Different Levels of Experience}
    \label{fig:MotivatingExample}
    \Description[ME]{Motivating Example}
\end{figure}

Code reviews can be conducted by a wide range of reviewers with varying levels of expertise, which may lead to diversity in the quality of reviews.
Prior studies found that reviewers' experience and expertise are often associated with the issue types identified in code reviews~\cite{kononenko2016code} and the level of usefulness~\cite{turzo2024}.
Inexperienced reviewers (who have reviewed and authored few code changes) often focus on trivial issues like visual representation (e.g., Figure~\ref{fig:MotivatingExample}, Example 2), which are considered less useful by developers~\cite{kononenko2016code,turzo2024}, or express confusion and uncertainty (e.g., Figure~\ref{fig:MotivatingExample}, Example 3), an anti-pattern associated with a lack of experience~\cite{antipatterns}. 
In contrast, experienced reviewers are more likely to identify critical functional issues, such as the missing validation check in Figure~\ref{fig:MotivatingExample}, Example 1, which demonstrates the value of deeper code understanding~\cite{kononenko2016code}.

As code reviews may reflect the reviewers' insights drawn from their prior software development and code review experiences, we hypothesise that the quality of generated reviews can be improved by aligning the language model with experienced reviewers' perspectives.
To achieve this alignment, we propose a suite of experience-aware training methods that utilise the reviewers' past authoring and reviewing experiences as signals for review quality.
Specifically, we propose a novel method called \textbf{experience-aware loss functions (ELF)}, which assigns a weight to the model's loss function that is proportional to the reviewer's experience in the project~\cite{aco,rso}.
In this way, comments made by experienced reviewers yield more influence over the model's behaviour.

Through both quantitative and qualitative evaluation, we found that ELF demonstrated stronger capability in generating high-quality reviews compared to past methods.
In terms of accuracy, ELF achieved the highest increase over CodeReviewer in terms of BLEU-4 (+5\%) and applicable comments generated (+29\%). 
In terms of informativeness, ELF exhibited the highest increase in suggestions over CodeReviewer (+56\%), whilst maintaining a similar improvement compared to experience-aware loss functions in terms of confused questions (-71\%) and explanations generated (+125\%).
Regarding the types of generated reviews, we found that ELF demonstrated the highest increase over CodeReviewer in terms of functional faults detected (+129\%) and evolvability issues identified (+21\%).
Amongst the different configurations of ELF, we found that considering both authoring and reviewing experience separately at the package level yielded the most improvement, however the complementary nature of the different granularities should also not be overlooked.

This study extends our short paper~\cite{Lin2024}, published at MSR 2024 (The International Conference on Mining Software Repositories), by introducing a new experience-aware technique along with deeper analysis to explore how different ways of estimating and weighing reviewers' experiences affect model-generated reviews.
To recapitulate, the initial work~\cite{Lin2024} introduced experience-aware oversampling, which aims to improve code review comment generation models by oversampling experienced reviewer's comments during model training. 
This preliminary method identified experienced reviewers by utilising the traditional 5\% ownership threshold rule~\cite{aco,rso} from both code reviewing and authoring perspectives at the repository level.
In this extension, our experimentation expands upon the initial work along two main vectors. 
Firstly, we expand the calculation of the ownership metrics to consider finer granularities; that is, we consider subsystem and package-level ownership in addition to repository-level ownership that was originally considered in the short paper.
The intention is to better reflect reviewers' specialised experiences within the project.
Secondly, we introduce experience-aware loss functions (ELF) to replace the previously proposed experience-aware oversampling method; we do this for two reasons: 1) to mitigate the sensitivity to arbitrarily selected major ownership thresholds that may not generalise across projects, and 2) to circumvent the computationally costly search for optimal upsampling rates.
We compare 12 new models derived from ELF in addition to the three previously proposed experience-aware oversampling strategies from the short paper (calculated at repository level only) and the original CodeReviewer model in terms of accuracy, informativeness, and issue types discussed.

The main contributions of this paper are:
\begin{itemize}
  \item A suite of experience-aware training methods for improving code review comment generation
  \item An analysis of emergent behaviours of code review comment generation models after experience-aware training
  \item Two large scale datasets containing commit and pull-request histories for 826 of the top GitHub repositories 
  \item An augmented version of CodeReviewer's dataset that is tagged with six different ownership metrics
\end{itemize}

\section{Background \& Related Work}
In this section, we discuss related work from three main areas: automated code reviews (2.1), review comment generation (2.2), and experienced reviewers and code review quality (2.3).
We focus on these aspects as we seek to improve review comment generation by leveraging reviewer experience as a signal for identifying higher quality reviews during model training.

\subsection{Automated Code Reviews}
The recent success of deep learning based language models in the field of natural language processing~\cite{vaswani2017attention,BERT,gpt,raffel2020exploring} has inspired a plethora of research on their application to the software engineering domain~\cite{lu2021codexglue,zeng_extensive,simplify}. This transition is seemingly natural, as both programming languages and natural languages are used to form sequential texts that conform to syntax and grammar, whilst expressing human logic in their semantics.
As a result, many studies have demonstrated that language models can achieve promising levels of performance in real-world program understanding and generation tasks~\cite{zeng_extensive} --- such as vulnerability detection~\cite{linevul}, repair~\cite{vulrepair,outofsight,vitrepair} and bug fixing~\cite{tufanobugfix,cure,jin2023inferfix} --- suggesting the existence of exploitable naturalness properties~\cite{ray2016naturalness,hindle2016naturalness,allamanis2018survey} within human written software.
Likewise, the field of automated code reviews also operates under the assumption that review comments are often repetitive and predictable, making the problem a suitable candidate for language modelling.

In its current form, automated code reviewing can be subdivided into a sequence of three tasks: code change quality estimation~\cite{codereviewer,hellendoorn2021towards}, review comment generation~\cite{codereviewer, tufano2022using, li2022auger,Lin2024}, and code refinement~\cite{codereviewer,tufano2019learning,tufano2021towards,tufano2022using,lin2023towards,li2023codeeditor,lin2023cct5,thongtanunam2022autotransform,pornprasit2023d}.
Firstly, code change quality estimation requires the model to determine whether a code change submitted by a developer requires code review.
This can be defined as a ($H_{pre} \rightarrow revise?$) binary classification task~\cite{codereviewer}, where $H_{pre}$ is the version of the code hunk submitted for review and $revise?$ is the binary decision output of whether the code change needs review.
If it is determined that the code change needs review, review comment generation is subsequently performed.
The review comment generation task (the subject of this study) requires the model to generate a natural language comment that can help guide the developer to improve the submitted piece of code, just as a human reviewer would. 
More formally, this can be formulated as a ($H_{pre} \rightarrow R_{nl}$) neural machine translation task~\cite{codereviewer}, where $R_{nl}$ is the natural language review comment.
Finally, code refinement requires the language model to address the review comment $R_{nl}$, by revising $H_{pre}$ to the final improved version of the code hunk $H_{post}$, ready to be merged into the code base.
This last task can be formulated as a ($H_{pre},R_{nl} \rightarrow H_{post}$) bimodal input translation problem~\cite{codereviewer}.

Whilst it has been proven that underlying statistical properties of code reviews are learnable, publicly available state-of-the-art techniques such as LLaMA~\cite{lu2023llama} and ChatGPT~\cite{guo2024exploring,tufano2024code} still struggle to perform well on the given test sets.
In contrast, the code review models trained on closed-source software projects~\cite{vijayvergiya2024ai,didact} have reported high adoption rates.
A likely reason being the data was collected from environments with experienced software engineers, strict code review cultures, and well-managed version control data, demonstrating that these code review models are performing well in practice when trained with high-quality examples.
However, such data of closed-source projects are limited, resulting in the need for models to rely on data from open-source platforms such as GitHub, where the code review standards may vary widely.
Our study focuses on alleviating this issue for the task of review comment generation, by leveraging reviewer experience as a proxy for identifying high quality code reviews in the open source.

\subsection{Review Comment Generation}
Review comment generation (the subject of this study) represents the most challenging task in the attempt to automate code reviews, as the language model is required to infer the single ground truth comment from a vast space of potential solutions using only a small window of code.
Tufano et al.~\cite{tufano2022using} have demonstrated the potential of using the text-to-text transfer transformer (T5)~\cite{raffel2020exploring} to generate code reviews, when pre-training the model with general code and natural language corpora related to the software engineering domain.
Yet, compared to code refinement, the model showed far lower performance in terms of text matching metrics for review comment generation; however, upon their manual inspection, it was revealed that many generated comments were in fact either semantically equivalent to the ground truth or a valid alternative solution.
In this study, we conduct manual evaluation to additionally account for these cases.

Whilst Tufano et al.~\cite{tufano2022using} conducted pre-training on more general software engineering corpora, Li et al.~\cite{auger} found success in jointly pre-training on code functions and their attached reviews.
Simultaneously, CodeReviewer~\cite{codereviewer} showed significant performance leaps by training on code review specific pre-training tasks as opposed to the generic masked language modelling technique used in prior works~\cite{tufano2022using,auger}.
This success was enabled by their large scale multilingual GitHub code review dataset, which is now widely employed as the benchmark dataset~\cite{Lin2024,sghaier2024improving,lu2023llama,lin2023cct5}.

More recently, Lu et al.~\cite{lu2023llama} discovered that generic LLMs such as LLaMA~\cite{touvron2023llama} were also capable of review comment generation by parameter-efficient fine-tuning on only 8.4 million parameters.
Taking a different approach, Sghaier and Sahraoui~\cite{sghaier2024improving} explored the effect of cross-task knowledge distillation, by jointly learning the successive tasks of review comment generation and code refinement together with two symbiotic models.

Whilst past research efforts have focused on improving review comment generation using various machine learning techniques, all of these approaches still assume the quality of comments across their open-source datasets are equivalent i.e., their approaches allow all data to contribute uniformly to the model's learning process. 
This can be suboptimal as the models will equally attempt to imitate the behaviour found in low quality code reviews, as those of higher quality, which can reduce the potential usefulness of the resulting tool, as evidenced by our initial results~\cite{Lin2024}.
In contrast, our current study proposes a novel method that forces more attention on potentially higher quality examples, such that they yield stronger influence over model behaviour, which may in turn increase the usefulness of the tool.

\subsection{Experienced Reviewers and Code Review Quality}
Detailed knowledge about program elements is usually retained by developers who frequently work with the code base~\cite{fritz2007}.
In the context of software quality, code ownership has often been a determinant of software defect proneness, where ownership ratios are inversely related to defect occurrences~\cite{aco,rso,hata2012bug}.
A study of implicated code has demonstrated that lower file level ownership tended to characterise the developers responsible for buggy lines~\cite{file_aco}, thus highlighting the importance of specialised knowledge.
Concurrently, experienced developers are often assigned to fix complex bugs due to their expertise~\cite{dormantbugs}, which recursively deepens their knowledge of potential software quality issues.
Given that code reviewers are a sub-population of developers, one would naturally intuit that their software development experience also has an impact on their effectiveness in reviewing code.

Thongtanunam et al.~\cite{rso} studied the relationship between software defects and code ownership of reviewers at the module level, concluding that modules that were reviewed by developers who lack both code authoring and reviewing expertise were more likely to be defect-prone.
Whilst studying the role of people and participation in code review quality, Kononenko et al.~\cite{peopleparticipation} also discovered corroborating evidence of the inverse relationship between reviewer experience and the presence of bugs, thus indicating that code review quality indeed varies depending on the individual who conducts it.
A survey in the open-source Mozilla core project revealed that developers were in close to uniform agreement regarding reviewer experience being a factor that influences code review quality~\cite{kononenko2016code}.
The developers reasoned that domain knowledge is crucial for properly evaluating a change, as superficial reviews arise from a lack of familiarity with the code base.
From an industry perspective, findings from Microsoft demonstrated that having prior experience with a file under review has a noticeable effect on the usefulness of the reviews provided~\cite{bosu2015}.
Whilst this line of research has studied the notion of useful comments coarsely defined by their ability to trigger a code change~\cite{bosu2015,rahman2017predicting}, our study focuses on a more fine-grained view of review quality, scrutinising the variation in quality between change triggering comments.

Past work have proposed to automatically evaluate code review quality by directly classifying review comments, however, they often do not cover detailed technical concerns~\cite{bosu2015, rahman2017predicting, evacrc, hasan2021}, struggle to reach reliable accuracy~\cite{turzo_comment_clf}, or are privately developed for specific software environments~\cite{evacrc}.
For example, Bosu et al.~\cite{bosu2015} and Rahman et al.~\cite{rahman2017predicting} framed the problem as a binary classification task, using classical machine learning models to identify useful comments i.e., change triggering comments.
On the other hand, Yang et al.~\cite{evacrc} proposed to assess comments based on linguistic attributes i.e., emotion, question, evaluation, suggestion, using a separate binary classifier for each attribute.
Although these types of classifications provide insight into communication dynamics and reviewer intent, they do not capture technical concerns of code changes under review, as both comments addressing critical issues regarding functional defects and trivial issues (e.g., visual representation) will trigger changes, evaluate code weaknesses and provide suggestions.
Different to this line of work, other studies have investigated code review quality from the perspective of issue categories.
Developers at Samsung~\cite{hasan2021} have expressed that reviews which identify defects, missing validations, performance optimisation opportunities, logical mistakes, etc. are useful, whilst visual representation issues that can be identified by static analysis tools are not useful.
Coinciding with these results, developers from the OpenDev project~\cite{turzo2024} rated reviews that discuss functional defects and validation issues as the most useful type of reviews, whilst those that address visual representation are rated as the least useful.
Based on this spectrum of comment usefulness ratings, a reviewer's prior coding experience was found to have the largest positive impact on comment usefulness, whilst their authorship/reviewership of the specific file under review had no significant association.
As the positive relationship between reviewer experience and code review quality has continuously resurfaced across different software development environments, and has shown promising results in our initial study~\cite{Lin2024}, we hypothesise that this attribute can directly serve as a general review quality signal for enhancing automated code review models in the absence of detailed and reliable review comment classifiers.

\section{Experience-Aware Training Methods}
Different to past approaches~\cite{codereviewer,auger}, which have mainly focused on the effectiveness of deep learning based language modelling techniques, our approach is the first to scrutinise the variation in quality amongst open-source code reviews.
As prior studies~\cite{turzo2024,hasan2021,kononenko2016code} found that reviewer experience is positively correlated with review quality, we developed two experience-aware model training methods to improve the quality of model generated code reviews.
This section presents our proposed experience-aware training methods.
In particular, we describe reviewer experience heuristics that are used to reflect a reviewer's software development experiences (3.1), our previously proposed experience-aware oversampling method (3.2), and our newly proposed experience-aware loss function method (3.3).

\subsection{Reviewer Experience Heuristics}
We measure reviewers' experience by calculating traditional code ownership metrics based on reviewers' past activity as a code reviewer~\cite{rso} and as a code author~\cite{aco}.
We consider three granularity levels of software systems: repository, subsystem, and package~\cite{chrev}.
Ownership metrics at each granularity level represent different levels of knowledge coverage, where the repository level reflects general knowledge within a repository and the package level reflects knowledge of a particular component.
We measure experience at different levels of granularity because some reviewers have ownership over specific components in the software system~\cite{chrev,revfinder}, whilst other reviewers are maintainers that oversee the whole project at a macro level~\cite{maintainer}.

For authoring experience, \textbf{Authoring Code Ownership (ACO)}~\cite{aco} represents the share of overall code contributions attributable to an individual.
It represents a developer's experience and coverage on a piece of software gained from hands on coding.
We formulate ACO as follows:

\begin{equation}\label{(1)}
ACO(D,G) = \frac{\alpha(D,G)}{C(G)},~~~G \in \{Repository, Subsystem, Package\}
\end{equation}

\noindent where $\alpha(D,G)$ is the number of commits in which reviewer $D$ has contributed to the software at the targeted granularity $G \in \{Repository, Subsystem, Package\}$ and $C(G)$ is the total commits at that granularity.
A higher ACO ratio indicates more experience as a code author for the targeted granularity in the system, and vice versa.
Whilst other studies~\cite{meng_acc,truck_factor} explore authorship at a line-level fidelity i.e., git blame, the scale of our study is orders of magnitude larger, rendering this calculation infeasible.
A recent study also found that commit-based ownership shares a stronger relationship with software quality than line-based ownership~\cite{thongtanunam2024}.
Thus, we resort to commit level calculations, which is also widely considered as a reasonable approximation~\cite{aco,rso,mcintosh_coverage}.

For a reviewing experience, \textbf{Review-Specific Ownership (RSO)~\cite{rso}} represents the share of overall code reviews attributable to an individual.
It represents a developer's experience and coverage of a part of software systems gained from reviewing code.
We formulate RSO as follows:

\begin{equation}\label{(2)}
RSO(D,G) = \frac{r(D,G)}{\rho(G)},~~G \in \{Repository, Subsystem, Package\}
\end{equation}

\noindent where $r(D,G)$ is the number of closed pull requests in which reviewer $D$ has reviewed (i.e., commented at least once) for a given granularity $G \in \{Repository, Subsystem, Package\}$ and $\rho(G)$ is the total number of closed pull requests for that given granularity.
A higher RSO ratio indicates more experience as a code reviewer for the chosen granularity in the system, and vice versa.

The three levels of granularity are determined as follows.

\begin{enumerate}
\item[\ding{118}] \textbf{Repository level} is the most coarsely grained, where authoring and reviewing activities are considered in terms of the entire repository of the project under development.
Specifically, all commits and reviewed pull requests mined from a repository are considered at this level. 
Ownership at the repository level represents general experience covering the entire project.

\item[\ding{118}] \textbf{Subsystem level} is represented by the top level of file directories in the repository~\cite{chrev}.
For example, the file \textit{\textbf{arch}/arm64/kernel/module.c} is considered to sit within the \textit{\textbf{arch}} subsystem.
A commit or reviewed pull request belongs to a subsystem if at least one of the changed files resides within that directory.
Ownership at the subsystem level represents coverage over a particular top-level component of the system.
Whilst we use the term ``subsystem'' to refer to the top-level directories in the source code hierarchy, we recognize that this view may not match an experienced developer's mental model of the software architecture~\cite{storey1999cognitive,murphy1995software,kruchten19954+}.  
However, we consider that this model is a reasonable proxy: the source hierarchy is known to all developers of the project and hence has currency to them.
In addition, the evaluation will be based on 826 GitHub repositories in our study; thus, devising a specialised architectural model for each repository is impractical and the models would be peculiar to our experiences rather than those of the project developers.
\item[\ding{118}] \textbf{Package level} is represented by the immediate folder that contains the target file~\cite{chrev}.
For example, the file \textit{\textbf{arch/arm64/kernel}/module.c} is considered to sit within the \textit{\textbf{arch/arm64/kernel}} package.
A commit or reviewed pull request belongs to a package if at least one of the changed files resides within that directory.
Ownership at the package level represents coverage over a direct set of co-located files.
Similarly, our measurement of a package is only an approximation, it does not reflect the ground truth system architecture.
The term ``package'' here is defined in terms of the hierarchical view of file locations, not to be confused with programming language constructs such as Java packages or C++ namespaces. 
\end{enumerate}

\subsection{Experience-Aware Oversampling}
Experience-aware oversampling was designed to over represent experienced reviewers' code reviews during training, such that their perspectives yield more influence over the model's behaviour.
We previously introduced the experience-aware oversampling method as a preliminary test of the concept that automated code review models could conform to the perspectives of experienced reviewers, thus improving the quality of generated reviews~\cite{Lin2024}.
This method utilised the traditional 5\% ownership threshold rule~\cite{aco,rso} to target three specific sub-populations of the dataset, resulting in three separate models.
The first group were major authors $MA$ $(ACO\geq5\%)$, the second group were major reviewers $MR$ $(RSO\geq5\%)$, and the last group were major reviewers \emph{and} major authors $MRMA$ $(ACO\geq5\% \phantom{.} and \phantom{.} RSO\geq5\%)$.
Each model was trained by oversampling one of these groups with an oversampling rate of 400\%. 
A sizable oversampling rate was chosen with the intention to elicit strong effects.

\subsection{Experience-Aware Loss Functions}
In this study, we introduce experience-aware loss functions (ELF), which uses the reviewers' assigned ownership ratios directly as loss function weights during training, such that their code reviews yield stronger influence over the model's behaviour.
Specifically, we augment the original negative log-likelihood loss for review comment generation~\cite{codereviewer} with an additional experience embedded weight term $\omega$~\cite{wang_hem,li_eem,wang_sem,dong_w_cel,vish_e_loss}.
This weight term dynamically accounts for the actual ownership values in their continuous form which both removes the need for the major ownership threshold assumption and the need for tuning an upsampling rate.
Furthermore, utilizing the ownership values in their original form retains continuous information and allows the model to capture the intricate differences between the examples.
We formulate the experience-aware loss function as follows:

\begin{equation}\label{(3)}
\mathcal{L}_{RCG} = \omega\sum_{t=1}^{k}-\log P(w_t|c,w_<t) 
\end{equation}

\noindent where $c$ is the submitted code change, $w_t$ is the current comment token, $w_<t$ are the comment tokens generated so far, and $k$ is the sequence length.
We formulate four weighting strategies to embed the experience values of ACO and RSO dimensions into the weight term $\omega$.

\begin{enumerate}
\item[\ding{118}] $\omega_{aco} = e^{1+aco}$ takes only authoring experience into consideration, speculating that reviewers who are prolific in coding provide higher quality code reviews.
Intuitively, the higher the ACO of the reviewer who wrote the comment, the heavier the penalty (i.e., the loss value is amplified by the weight term $\omega$ ) to the model for an incorrect prediction.

\item[\ding{118}] $\omega_{rso} = e^{1+rso}$ takes only reviewing experience into consideration, speculating that reviewers who are prolific in reviewing pull requests provide higher quality code reviews.
Intuitively, the higher the RSO of the reviewer behind the comment, the heavier the penalty to the model for an incorrect prediction.

\item[\ding{118}] $\omega_{avg} = e^{1+\frac{rso+aco}{2}}$ considers both authoring and reviewing experience equally, speculating that reviewers who are prolific in both committing and reviewing code provide higher quality code reviews.
Intuitively, the higher the combined ACO and RSO of the reviewer behind the comment, the heavier the penalty to the model for an incorrect prediction.

\item[\ding{118}] $\omega_{max} = e^{1+max(rso,aco)}$ considers only the most representative experience type, speculating that reviewers who are prolific in either committing or reviewing code provide higher quality code reviews.
Intuitively, the higher the reviewer's max ownership, regardless of experience type, the heavier the penalty to the model for an incorrect prediction.
\end{enumerate}

Given that the model will be penalised differently based on the particular reviewer experience in focus, the weighted loss forces the model to align to their code review examples by affecting the direction of the gradient updates.
Since the variation between ownership values is numerically small, we take an exponential to create stronger separation effects~\cite{dong_w_cel,wei_e_loss,vish_e_loss}.
The four weights can be calculated at the three aforementioned granularities $G \in \{Repository, Subsystem, Package\}$, resulting in 12 different models.
For example, the model trained with $\omega_{aco-Repo}$ weighted $\mathcal{L}_{RCG}$ conforms to reviewers who have high authoring code ownership at the repository level, whilst the model trained with $\omega_{avg-Pkg}$ weighted $\mathcal{L}_{RCG}$ aligns to reviewers with high coverage in terms of both authoring and reviewing ownership at the package level.

\section{Experiment Setup}
In this section, we propose our research questions (4.1), provide a study overview (4.2), as well as describe our data preparation steps (4.3), fine-tuning implementation (4.4), evaluation metrics (4.5), and manual evaluation process (4.6).

\subsection{Research Questions}
In this work, we set out to investigate the effectiveness of experience-aware training methods on code review comment generation.
We formulate three research questions to evaluate the model performance in three main aspects, i.e., accuracy, informativeness, and issue types.

\textbf{(RQ1) What is the impact of ELF on the accuracy of code review comment generation models?} 
\\
\underline{\textit{Motivation:}} 
This RQ aims to evaluate the correctness of the generated comments against the ground truth, i.e., comments made by human reviewers.
We measure the accuracy of the ELF models in terms of textual similarity~\cite{bleu} and semantic equivalence~\cite{tufano2021towards,tufano2022using,auger} against the ground truth in the test set.
Additionally, to capture the diverse nature of all potential code reviews, we also assess their general applicability~\cite{tufano2021towards,tufano2022using,codereviewer} to the code change submission.
The textual similarity metric (BLEU-4), will be automatically assessed, whilst both semantic equivalence and applicability are manually evaluated.

\textbf{(RQ2) What is the impact of ELF on the informativeness of code review comment generation models?} 
\\
\underline{\textit{Motivation:}} 
Informativeness is one of the important characteristics of high-quality reviews as perceived by developers~\cite{hasan2021,kononenko2016code}.
Typically, the informativeness of a code review refers to whether they identify an issue~\cite{bosu2015}, prescribe a solution~\cite{hasan2021,codereviewer}, and provide a clear explanation for their rationale~\cite{kononenko2016code,turzo2024,trenches,rahman_example}. 
The ability to achieve all three criteria may vary depending on the reviewer's experience.
In this RQ, we set out to evaluate the quality of the code reviews generated by the ELF models in terms of feedback type, i.e. suggestions, concerns, confused questions and presence of explanation. 
Both feedback type and presence of explanation are metrics that involve manual evaluation.

\textbf{(RQ3) What issue types are discussed in the ELF models' code reviews?} 
\\
\underline{\textit{Motivation:}} 
Code reviews cover a wide variety of issues, including both functional and non-functional properties of the system~\cite{mantyala}.
Technical issues such as those related to defects and logic often require more in-depth knowledge of the system~\cite{kononenko2016code,turzo2024} as opposed to generic issues such as improving visual representation.
As developers seek more insightful code reviews~\cite{kononenko2016code,hasan2021,turzo2024,trenches}, it is crucial that the model is able to offer comments that reflect deeper issues.
For this RQ, we manually evaluate the issue types discussed in the ELF models' code reviews.

\begin{figure}
    \centering
    \includegraphics[width=1\textwidth]{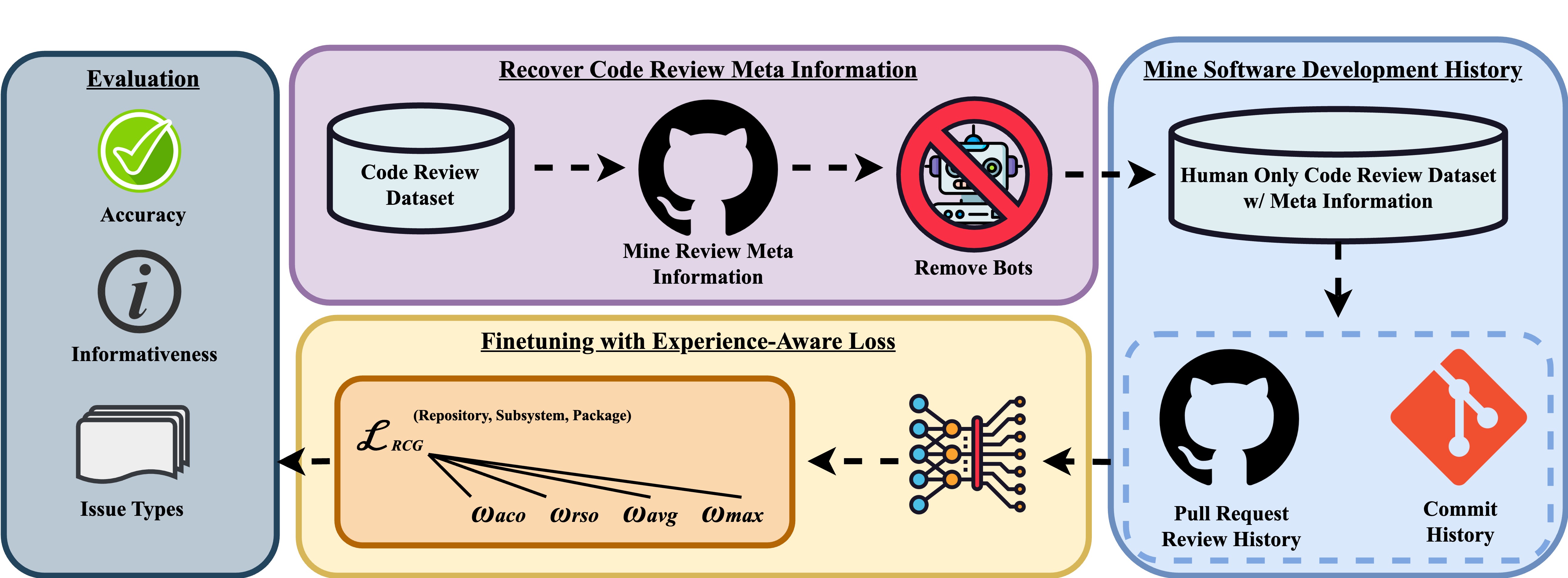}
    \caption{Overview of Our Experimental Design}
    \label{fig:EPD}
    \Description[EPD]{An Overview of Our Experimental Design}
\end{figure}

\subsection{Study Overview}
The overview of our experimental design is presented in Figure~\ref{fig:EPD}.
To answer all three RQs, we need to 1) construct a code review dataset tagged with ownership ratios to facilitate experience-aware fine-tuning, 2) fine-tune CodeReviewer's pre-trained checkpoint using each strategy i.e., ELF, experience-aware oversampling, original, and 3) evaluate all of the resulting models with respect to the metrics proposed for each of the RQs.
To evaluate the degree of impact each strategy has on accuracy, RQ1 employs three accuracy metrics i.e., BLEU-4, semantic equivalence and applicability.
Whilst BLEU-4 is an automatic text matching metric that can be applied to the entire test set, semantic equivalence and applicability measure accuracy based on the actual meaning being conveyed in the review, which requires human level understanding.
As a result, these two metrics are manually evaluated based on 100 random samples in the test set.
To evaluate the degree of impact each strategy has on informativeness, RQ2 employs two metrics i.e., feedback type, presence of explanation.
Since these two metrics also rely on human level understanding to determine the level of information provided in the review, they are manually evaluated as well.
Finally, to evaluate the degree of impact each strategy has on the issue types discussed, RQ3 requires manual evaluation to reliably categorise the content of the review.
The details of the experimental design are discussed below (Sections 4.3 - 4.6).

\subsection{Data Preparation}

\textbf{Dataset selection.}
We use the CodeReviewer dataset provided by Li et al.~\cite{codereviewer}, which is the benchmark dataset for the field of automated code reviews.
Specifically, we use the code refinement set for review comment generation, as their original comment generation dataset did not retain pull request IDs, making them untraceable.
The dataset represents a diverse set of software projects written in nine of the most popular programming languages on GitHub, including Python, Java, Go, C++, JavaScript, C, C\#, PHP, and Ruby.
The training set was derived from 519 repositories, representing a subset of the top 10k most starred projects that contained more than 2.5k pull requests.
The validation and test sets were derived from 307 repositories that contained between 1.5k and 2.5k pull requests.
The size of the original dataset was 150,406 for training, 13,103 for validation and 13,104 for testing.

\textbf{Meta information recovery.}
To retrieve meta information for each comment in the dataset, we used PyGithub\footnote{\label{PyGitHub}\url{https://github.com/PyGithub/PyGithub}} to retrieve the associated pull requests and identify the original comments using string matching.
For each comment, we extracted the username and ID of the reviewer and the timestamp of when the comment was posted.
Throughout this process, we identified 10,583 accounts who wrote reviews in the training set and 2,763 accounts in validation and test set.
The reviews covered a timeframe between 2011 and 2022.

\textbf{Preprocessing.}
Given that our focus is to build a review comment generation model that reflects the experienced perspective of human reviewers, we removed all bot accounts (e.g., CI bots, style checkers).
This was achieved through two common methods~\cite{bot_detection}: 1) identification by the ``bot'' suffix~\cite{bot_suffix}, and 2) identification by an established list of bots~\cite{botlist}.
The identified bot accounts were manually inspected for false positives, which were subsequently retained.
Finally, we discarded comments that provided code but without any natural language remarks.
More specifically, we are referring to the GitHub code suggestion functionality that can be found in comments denoted with the $```suggestion```$ format, where the reviewer's comment contains only a code edit of the original code hunk under review. 
This type of comment makes up approximately 5\% of the refinement dataset~\cite{Lin2024}, which caused severe mode collapse in the model when used for comment generation training.
Without removing the code-only comments, the model's ability to provide real natural language comments can be degraded by only copy-and-pasting entire chunks of the submitted code, whilst inflating its performance on text matching-based results, e.g., BLEU-4.
The final dataset included 141,259 examples for training, 12,406 for validation and 12,369 for testing.
Table~\ref{tab:data_transformation} presents statistics regarding the dataset filtering process.

\textbf{Mining software development history.}
We collected meta information on all commits and pull requests for each of the 826 GitHub repositories.
More specifically, for each commit, we retrieved the original author, a list of the changed files, and the timestamp using PyDriller~\cite{PyDriller} and for each pull request, we retrieved the GitHub users who left comments, a list of the changed files, and the timestamp using PyGithub\footref{PyGitHub}.
We observed that only some groups of developers used the GitHub review comment function, while others left reviews in the form of issue comments on the pull request.
Therefore, in addition to considering GitHub review comments such as in our previous work~\cite{Lin2024}, we also consider issue comments as they demonstrate code reviewing activity~\cite{perils}.
In terms of commit history, we omit merge commits as they do not represent actual code authoring activity.
Table~\ref{tab:data_transformation} presents statistics regarding the repository history mining process.

\textbf{Calculating ownership ratios.}
We calculate the ownership metrics with regard to each example (i.e., review comment) in the dataset.
Although each reviewer may provide multiple reviews, the ownership metrics are calculated independently for each review comment. The rationale is that a reviewer's ownership coverage may change throughout time, hence our method reflects a version of that reviewer distilled at the timestamp of the review comment.
The implementations are detailed in Algorithm~\ref{alg:aco_alg} for ACO calculation and Algorithm~\ref{alg:rso_alg} for RSO calculation, where $RC$ denotes a review comment, $D$ denotes a developer, and $G$ denotes the granularity level. 
Table~\ref{tab:data_transformation} presents statistics regarding the calculated ownership ratios.

\begin{table}[t]
\caption{Dataset Overview.}
\label{tab:data_transformation}
\begin{tabular}{l
>{\columncolor[HTML]{EFEFEF}}l 
>{\columncolor[HTML]{C0C0C0}}l 
>{\columncolor[HTML]{9B9B9B}}l }
                              & \textbf{Training}      & \textbf{Validation}    & \textbf{Test}     \\ \Xhline{3\arrayrulewidth}

\multicolumn{4}{c}{\textbf{Dataset Filtering (Reviews)}} \\ \Xhline{3\arrayrulewidth}
Original Dataset Size                                    & 150,406 & 13,103 & 13,104 \\ \hline
Untraceable        & - 618     & - 24     & - 41     \\
Generated by Bot                 & - 1,207   & - 41     & - 55     \\ 
No Natural Language Comment  & - 7,322   & - 632    & - 639    \\ \hline
Final Dataset Size                                       & 141,259 & 12,406 & 12,369 \\ \Xhline{3\arrayrulewidth}

\multicolumn{4}{c}{\textbf{Mining Repository History}} \\ \Xhline{3\arrayrulewidth}
\# Repositories   & 519   & 300    & 300    \\ \hline
\# Reviewer Accounts                                & 10,583  & 2,148  & 2,125  \\
\# Bot Accounts                          & 9       & 3      & 4      \\ \hline
\# Past Commits  & 8,294,486   & 2,945,639    & 2,944,569    \\
\# Past Closed Pull Requests  & 3,478,749   & 606,148    & 605,649    \\ \Xhline{3\arrayrulewidth}

\multicolumn{4}{c}{\textbf{Ownership Statistics ($\mu/\sigma$)}} \\ \Xhline{3\arrayrulewidth}
RSO (Repository)                & 0.21 / 0.21       & 0.31 / 0.24      & 0.31 / 0.24     \\ 
ACO (Repository)               & 0.08 / 0.13       & 0.15 / 0.21      & 0.15 / 0.21      \\ \hline
RSO (Subsystem)                & 0.25 / 0.23        & 0.35 / 0.26      & 0.35 / 0.26      \\ 
ACO (Subsystem)                & 0.1\phantom{0} / 0.14       & 0.17 / 0.22      & 0.17 / 0.22      \\ \hline
RSO (Package)                 & 0.31 / 0.27       & 0.39 / 0.29      & 0.38 / 0.29      \\ 
ACO (Package)                & 0.12 / 0.19       & 0.18 / 0.24       & 0.18 / 0.24      \\ \Xhline{3\arrayrulewidth}

\end{tabular}
\end{table}

\begin{algorithm}
\caption{Implementation of Authoring Code Ownership (ACO)}
\label{alg:aco_alg}
\begin{algorithmic}[1]
\Procedure{ACO}{$D,G,RC$} \Comment{ACO for one $RC$ example}
    \State $\alpha(D,G) \gets 0$
    \State $C(G) \gets 0$
    \ForAll{commit $\in G$}
    \State $\tau1 \gets$ $RC$ timestamp 
    \State $\tau2 \gets$ commit timestamp
    \If{$\tau2<\tau1$}
    $C(G) \gets C(G) + 1$ \Comment{\# commits in $G$ prior to $RC$}
    \If{$D\equiv$ commit author}
    $\alpha(D,G) \gets \alpha(D,G) + 1$ \Comment{\# commits in $G$ by $D$ prior to $RC$}
    \EndIf
    \EndIf
    \EndFor
    \State \textbf{return} $\alpha(D,G)/C(G)$ \Comment{ratio of commits in $G$ by $D$ over total commits in $G$ at $\tau1$}
\EndProcedure
\end{algorithmic}
\end{algorithm}

\begin{algorithm}
\caption{Implementation of Review-Specific Ownership (RSO)}
\label{alg:rso_alg}
\begin{algorithmic}[1]
\Procedure{RSO}{$D,G,RC$} \Comment{RSO for one $RC$ example}
    \State $r(D,G) \gets 0$
    \State $\rho(G) \gets 0$
    \ForAll{closed PR $\in G$}
    \State $\tau1 \gets$ $RC$ timestamp 
    \State $\tau2 \gets$ closed PR timestamp
    \State $\theta \gets$ \textbf{Set}(reviewers of closed PR)
    \If{$\tau2<\tau1$}
    $\rho(G) \gets \rho(G) + 1$ \Comment{\# closed PRs in $G$ prior to $RC$}
    \If{$D\in \theta$}
    $r(D,G) \gets r(D,G) + 1$ \Comment{\# closed PRs in $G$ reviewed by $D$ prior to $RC$}
    \EndIf
    \EndIf
    \EndFor
    \State \textbf{return} $r(D,G)/\rho(G)$ \Comment{ratio of closed PRs in $G$ reviewed by $D$ over total closed PRs in $G$ at $\tau1$}
\EndProcedure
\end{algorithmic}
\end{algorithm}

\subsection{Model Fine-tuning}
Following CodeReviewer~\cite{codereviewer}, we fine-tuned the models with a learning rate of 3$e^{-4}$ using the AdamW optimizer. 
The training was set for 30 epochs at a batch size of 72.
A beam search width of 10 was used during inference.
Our hardware consisted of a single server with 32 CPUs, 256GB RAM and four NVIDIA A100-80GB GPUs.
For fair comparison with the experience-aware models, we fine-tuned CodeReviewer on our filtered dataset.
To incorporate our experience-aware loss functions and the previous experience-aware oversampling method, we altered the original Python scripts provided by Li et al.~\cite{codereviewer}, which were implemented with Pytorch\footnote{https://pytorch.org/} and HuggingFace Transformers\footnote{https://huggingface.co/}.
For a fair comparison, we also re-implemented experience-aware oversampling using our new ownership values, which were calculated based on our newly mined repository histories (i.e., omitting merge commits, including issue comments).
In total, we trained 16 models, including 12 ELF models, three experience-aware oversampling models, and one original CodeReviewer model.

\subsection{Evaluation Metrics}
Our evaluation consists of one automatic text matching metric and five manual evaluation tasks.
Automatic text matching is measured on the entire test set, whilst the manual evaluation tasks are completed on a random sample of 100, achieving a 95\% confidence level with 10\% margin of error.

\begin{enumerate}
\item[\ding{118}] 
\textbf{BLEU-4 (RQ1): } To align with past research~\cite{codereviewer}, we adopt the established Bilingual Evaluation Understudy metric~\cite{bleu} with up to 4-gram matching to benchmark the new approaches in terms of accuracy against the test set.
We use the exact implementation provided by Li et al.~\cite{codereviewer}, with stop words removed from the comments.

\item[\ding{118}] \textbf{Semantic Equivalence (RQ1): } 
This manual assessment of accuracy evaluates whether the model generated comments possess the same intentions as the ground truth code reviews in the test set~\cite{tufano2021towards,tufano2022using,auger}.
This metric disregards the degree of textual overlap, since the same intention can be expressed in various different ways.
For example, in Figure~\ref{fig:ME1}, despite being worded differently, candidate A is considered semantically equivalent to the ground truth as the main intention of both comments is to change the error to a log warning.

\item[\ding{118}] \textbf{Applicability (RQ1): }
This manual assessment of accuracy evaluates whether model-generated comments provide applicable reviews given the context of the submitted code change~\cite{tufano2021towards,tufano2022using,codereviewer}.
Since ground truth code reviews reflect only a subset of many possible issues with a particular piece of code, this evaluation captures the models' general ability to provide relevant code reviews.
For example, in Figure~\ref{fig:ME1}, comment candidate B is not considered to be semantically equivalent to the ground truth; however, it is still applicable to the code change as the developer can make subsequent code improvements to address the comment.
By definition, all comments that are semantically equivalent to the ground truth are also applicable.

\begin{figure}
    \centering
    \includegraphics[width=0.65\textwidth]{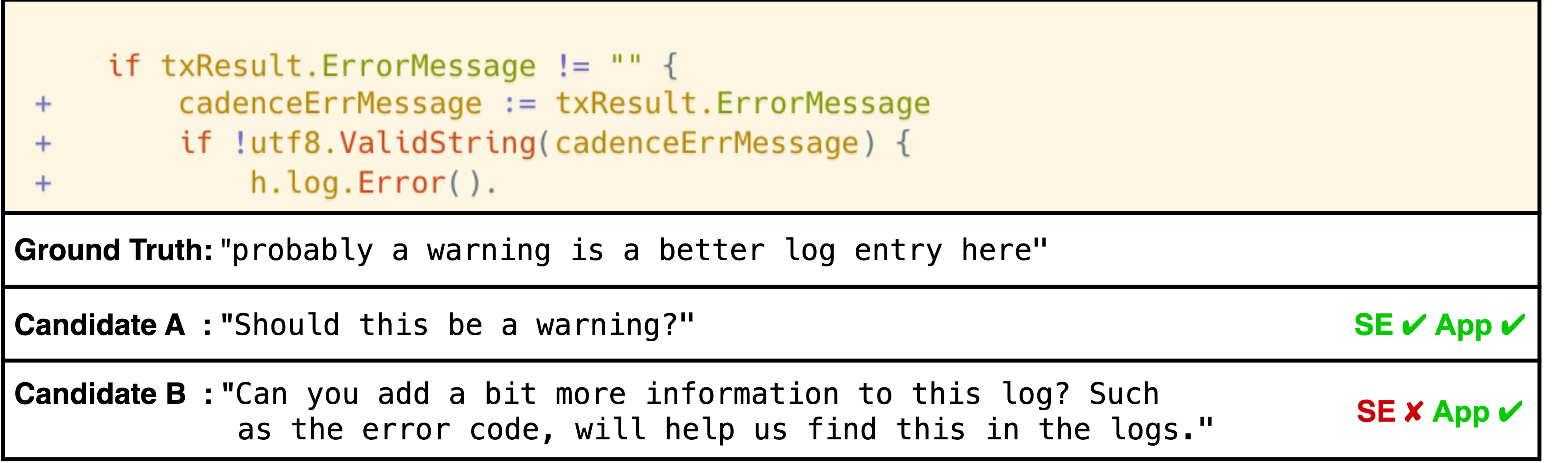}
    \caption{Examples of Semantically Equivalent (SE) and Applicable (App) Comments}
    \label{fig:ME1}
    \Description[ME]{Manual Example}
\end{figure}

\begin{figure}
    \centering
    \includegraphics[width=1\textwidth]{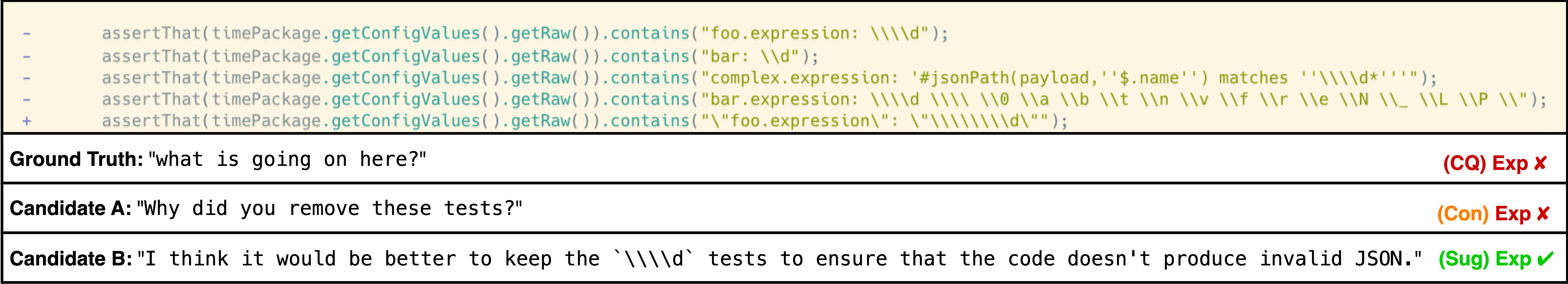}
    \caption{Examples of Suggestion (Sug) with Explanation (Exp), Concern (Con) and Confused Question (CQ)}
    \label{fig:ME2}
    \Description[ME]{Manual Example}
\end{figure}

\item[\ding{118}] \textbf{Feedback Type (RQ2): }
To measure informativeness, this manual assessment categorises the code reviews into three distinct feedback types.
We present them in order of most to least informative:
\begin{itemize} 
            \item \textit{Suggestion}~\cite{codereviewer,bosu2015,hasan2021} --- Not only is an issue identified, but a solution is also proposed to address the issue.
            \item \textit{Concern}~\cite{codereviewer,hasan2021} --- An issue is identified or doubt is raised; however, no solution is provided.
            \item \textit{Confused Question}~\cite{antipatterns} --- The comment demonstrates an inability to comprehend the code change.
\end{itemize}
In Figure~\ref{fig:ME2}, the most uninformative feedback type is exemplified in the ground truth, which is a confused question that demonstrates an inability to understand the code change.
Comment candidate A will be categorised as a concern because only a concern is raised, but no further solution is provided.
Comment candidate B is considered as providing a suggestion because it recognises that the assert statements should not be removed and directly suggests to "keep the `\textit{\textbackslash\textbackslash\textbackslash d' tests}".

\item[\ding{118}] \textbf{Presence of Explanation (RQ2): }
This manual evaluation is a binary measure of whether the comment expresses their rationale.
A comment that explains itself is considered to be more informative~\cite{kononenko2016code,turzo2024,trenches,rahman_example}.
For example, in Figure~\ref{fig:ME2}, we consider that comment candidate B has expressed their rationale as it describes why "`\textit{\textbackslash\textbackslash\textbackslash d}' tests" should be kept \textit{"to ensure that the code doesn't produce invalid JS0N"}.
On the contrary, both the ground truth and candidate A do not provide a rationale for their concerns or questions.

\item[\ding{118}] \textbf{Comment Category (RQ3): }
This manual evaluation categorises the suggestions and concerns based on 15 established issue categories developed by past work~\cite{beller,bosu2015,mantyala,turzo2024}.
These categories are broadly segmented into three large classes, covering functional issues, evolvability issues, and discussions.
Functional issues are defects that can cause system failures at execution time, whilst evolvability issues are non-functional issues that affect the compliance, maintainability and understandability of the code.
Discussions are dialogues that invoke thought regarding design directions and implementation choices.
We only label comment category for comments previously annotated as applicable, as such, we do not consider the praise and false positive categories like in prior work~\cite{turzo2024}.
Comments that do not fit into the taxonomy are categorised as other.
The full list of code review comment categories and their respective descriptions are detailed in Table~\ref{table:category_defs}.
\end{enumerate}

\begin{table}[]
\caption{Code Review Comment Categories}
\label{table:category_defs}
\begin{tabular}{|ll|l|}
\hline
\multicolumn{1}{|l|}{Group} & Category & Description \\ \hline
\multicolumn{1}{|l|}{\multirow{7}{*}{Functional}} & Functional Defect & \begin{tabular}[c]{@{}l@{}}A functionality is missing or implemented incorrectly, which often \\ requires  additional code or larger modifications.\end{tabular} \\ \cline{2-3} 
\multicolumn{1}{|l|}{} & Validation & \begin{tabular}[c]{@{}l@{}}Issues with detecting an invalid value and issues related to data \\ sanitisation.\end{tabular} \\ \cline{2-3} 
\multicolumn{1}{|l|}{} & Logical & \begin{tabular}[c]{@{}l@{}}Issues with comparison operations, control flow, computations \\ and other types of logical errors\end{tabular} \\ \cline{2-3} 
\multicolumn{1}{|l|}{} & Interface & \begin{tabular}[c]{@{}l@{}}Issues when interacting with other parts of the software \\ e.g., existing code library, hardware device, database, operating system\end{tabular} \\ \cline{2-3} 
\multicolumn{1}{|l|}{} & Resource & \begin{tabular}[c]{@{}l@{}}Issues with the initialisation, manipulation and release of variables, \\ memory, files and database\end{tabular} \\ \cline{2-3} 
\multicolumn{1}{|l|}{} & Support & Issues related to support systems, libraries or their configurations \\ \cline{2-3} 
\multicolumn{1}{|l|}{} & Timing & Issues with incorrect thread synchronisation in shared resource settings \\ \hline
\multicolumn{1}{|l|}{\multirow{6}{*}{Evolvability}} & Solution Approach & Suggestions for alternate implementations e.g., algorithms, data structures \\ \cline{2-3} 
\multicolumn{1}{|l|}{} & Documentation & Suggestions to improve code comments or documentation \\ \cline{2-3} 
\multicolumn{1}{|l|}{} & Organisation of Code & \begin{tabular}[c]{@{}l@{}}Suggestions for structural refactoring \\ e.g., collapse hierarchy, extract super class, inline function\end{tabular} \\ \cline{2-3} 
\multicolumn{1}{|l|}{} & Alternate Output & \begin{tabular}[c]{@{}l@{}}Suggestions for improving error messages, toast messages, alerts \\ and the returned values of a function\end{tabular} \\ \cline{2-3} 
\multicolumn{1}{|l|}{} & Naming Convention & Suggestions for renaming software elements to comply with conventions \\ \cline{2-3} 
\multicolumn{1}{|l|}{} & Visual Representation & \begin{tabular}[c]{@{}l@{}}Suggestions for improving code readability \\ e.g., removing white spaces, blank lines, code rearrangements, indentation\end{tabular} \\ \hline
\multicolumn{1}{|l|}{\multirow{2}{*}{Discussion}} & Question & \begin{tabular}[c]{@{}l@{}}Questions to understand design and implementation choices\end{tabular} \\ \cline{2-3} 
\multicolumn{1}{|l|}{} & Design Discussion & \begin{tabular}[c]{@{}l@{}}Higher level discussions on design directions, design patterns \\ and software architecture\end{tabular} \\ \hline
\multicolumn{2}{|c|}{Other} & Comments that do not fit within the taxonomy \\ \hline
\multicolumn{3}{l}{\footnotesize The code review comment categories and their respective descriptions are adopted from past work~\cite{turzo2024,mantyala,beller}} \\
\end{tabular}
\end{table}

\subsection{Manual Evaluation Process}
To answer all three research questions, we rely on manually evaluated metrics i.e, semantic equivalence (RQ1), applicability (RQ1), feedback type (RQ2), presence of explanation (RQ2) and comment categories (RQ3).
We detail the evaluation process and discuss the inter-rater agreement for these five metrics below.

\textbf{Overall manual evaluation process.} The manual evaluation was conducted by the first and sixth authors.
Both annotators have previously been employed as software engineers and are currently pursuing a PhD in software engineering.
One has five years of software development experience, whilst the other has over 10 years.
The manual evaluation consisted of 8,000 annotations (100 generated comments $\times$ 16 models $\times$ 5 manual evaluation tasks).
The annotators were provided with a guideline including the definitions above, the submitted code hunk, the ground truth review comment, the generated reviews, and the post-review code refinement.
To mitigate incorrect annotations caused by a lack of familiarity with the software environments in the examples, the annotators may acquire more information from the original repositories and external resources such as Stack Overflow and official package/language documentation.
For each manual evaluation task (1,600 annotations for 100 generated comments $\times$ 16 models), the two annotators independently completed two rounds of annotations where the first round consisted of 300 generated comments and the second round consisted of 200 generated comments.
After each round, we measure inter-rater agreement and resolve all conflicts,
resulting in a refined annotation guideline that both annotators agreed on~\cite{baltes_doc}.
After reaching an agreement rate that we were satisfied with~\cite{techcontext} ($\geq$0.8 Cohen's kappa~\cite{cohen1960}), the first author annotated the remaining 1,100 comments independently.
Finally, the annotations were reviewed by the second and fifth authors.
Note that we omit information regarding which models generated the comments during the annotation process to eliminate a confirmation bias, i.e., avoiding the evaluation results favoring particular models.
Below, we summarise our annotation process for each evaluation task, including the inter-rater agreement achieved during the independent annotation rounds and the conflicting cases that were resolved.

\textbf{Semantic equivalence (RQ1).}
The annotators reached 89\% agreement with a Cohen's kappa of 0.66 (substantial agreement) in the first round; the second round reached a satisfactory agreement rate of 94\% with a Cohen's kappa of 0.82 (near perfect agreement).
The most common type of conflict arose from the cases where the generated comments 
have semantically equivalent intention as the ground truth but suggest incorrect implementation, and where the generated comments have ambiguous intentions which are open to multiple interpretations.
The comments with partial semantic equivalence were eventually labelled as not semantically equivalent as the suggested implementation would lead to the wrong fix.
For the comments with ambiguous intentions, they were labelled as semantically equivalent if the subsequent ground truth code change was in the potential action space that could be elicited by the generated comment.
Otherwise, they were labelled as not semantically equivalent.

\textbf{Applicability (RQ1).}
The annotators reached 83\% agreement with a Cohen's kappa of 0.67 (substantial agreement) in the first round; the second round reached a satisfactory agreement rate of 93\% with a Cohen's kappa of 0.86 (near perfect agreement).
The most common types of conflicts arose from cases where the generated comments required additional context and/or project knowledge to comprehend.
For example, a generated comment suggested to rename a method  "\textit{fc\_fit\_scheduler}" to \textit{BasicTrainScheduler}", which can be considered as applicable. 
However, this suggestion is not suitable for a Python project as Pascal case violates PEP8 guidelines.
These types of conflicts were resolved by consulting external resources in the context of the project.

\textbf{Feedback type (RQ2).}
The annotators reached 93\% agreement with a Cohen's kappa of 0.86 (near perfect agreement) in the first round.
Given that near perfect agreement was achieved, we omitted the second round and the first author performed the annotations on the remaining sampled comments.
The most common types of conflicts arose from cases where the comments raised concerns in a question form which were similar to suggestions in question form.
Code review suggestions are often provided in question form to be polite~\cite{intentions}, however, they can also be a request for confirmation.
For example, "\textit{Maybe close the `DeflaterOutputStream` here?}" and "\textit{Do we need to close the `DeflaterOutputStream` here?}" are similar comments, however the former is a polite suggestion that offers a solution, whilst the latter is a concern that requires other developers to clear their doubts.
The conflicts were resolved by considering comments with "\textit{Do we ...?}" questions as exhibiting uncertainty and therefore should be labelled as concern.

\textbf{Presence of explanation (RQ2).}
The annotators reached 99\% agreement with a Cohen's kappa of 0.98 (near perfect agreement) in the first round.
Given that near perfect agreement was achieved, the first author performed the annotations on the remaining sampled comments without the second round.
There was only one conflicting case where the comment asking "\textit{Isn't there a complete\_bipartite\_graph function?}" can be considered as a self-embedded rationale when viewed as a suggestion, and as lacking a rationale when viewed as a concern asking about the code change itself. 
This case was resolved by considering with the context of the feedback type and code change.

\textbf{Comment category (RQ3).} 
The annotators reached 88\% agreement with a Cohen's kappa of 0.86 (near perfect agreement) in the first round.
Given that near perfect agreement was achieved, we omitted this task from the second round.
The most common types of conflicts arose from the misallocation of variable declaration changes to the resource category.
This category includes variable initialisation changes, which distinctly focuses on problems related to resource allocation.
Since changes such as improving the declared variable type do not alter the external behaviour of the code, we categorise them as organisation of code a.k.a refactoring type code reviews instead.

\section{Preliminary Analysis on Code Review Comments and Reviewer Experience }

Before evaluating the models trained with ELF, we first investigate whether the nature of code reviews varies with reviewer experience in reality.
To investigate this, we compare the real distributions of informativeness and issue types discussed between the review comments of inexperienced and experienced reviewers.
To this end, we sample code reviews of reviewers with polar opposite degrees of experience from the training set.
We take extreme examples as there are no hard thresholds for determining general experience groups.
Since experience likely exists on a spectrum, we abstain from using thresholds on intermediary ownership values to group code reviews.
Thus, for this analysis, we consider inexperienced reviewers as those where all of their RSO and ACO values are two standard deviations below the mean or zero (RSO Repository = 0, ACO Repository = 0, RSO Subsystem = 0, ACO Subsystem = 0, RSO Package = 0, ACO Package = 0).
Conversely, we consider experienced reviewers as those where all of their RSO and ACO values are two standard deviations above the mean (RSO Repository $\geq$ 0.63, ACO Repository $\geq$ 0.34, RSO Subsystem $\geq$ 0.71, ACO Subsystem $\geq$ 0.38, RSO Package $\geq$ 0.85, ACO Package $\geq$ 0.5).
Based on these extreme thresholds, we extracted a pool of 1,031 reviews from inexperienced reviewers and 232 reviews from experienced reviewers.
We randomly sample 88 examples from both pools of reviews, achieving a 95\% confidence level with 10\% margin of error.
The first and sixth authors then manually annotated the sampled reviews together in terms of feedback type, presence of explanation and comment categories.
The reviewer demographic information was omitted to prevent confirmation bias.

\begin{figure}
    \centering
    \includegraphics[width=1\textwidth]{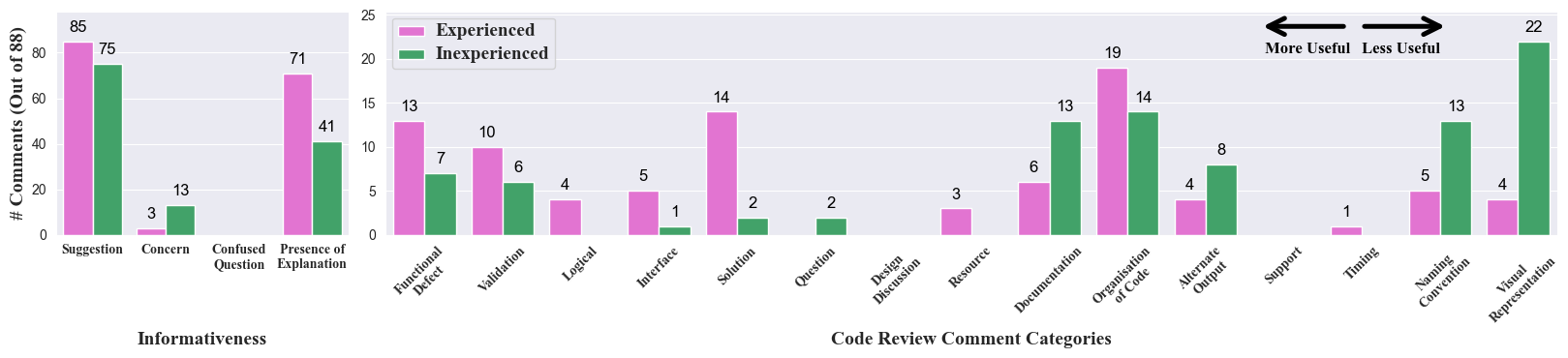}
    \caption{Comparing Informativeness and Comment Categories of Code Reviews by Experienced vs Inexperienced Reviewers}
    \label{fig:rd}
    \Description[rd]{Real Distribution}
\end{figure}

Figure~\ref{fig:rd} (left) shows the comparative results in terms of informativeness metrics.
We find that experienced reviewers provided more suggestions and explanations than inexperienced reviewers.
In particular, the difference in amount of explanations was substantial ($\Delta 34\%$), this phenomenon is within expectation as code reviews are often used by senior developers as a platform for knowledge transfer within a project~\cite{expectations}.
We find that explanations are prevalent when experienced reviewers are discussing complex issues e.g., deep functional defects, alternative implementation solutions.
Figure~\ref{fig:rd} (right) shows the comparative results in terms of comment categories.
These categories are ordered from most useful (left) to least useful (right) as rated by open source developers~\cite{turzo2024}.
We find that the code reviews from experienced developers tend to focus on more critical problems such as functional defects, validation issues, logical errors and incorrect usage of interfaces, whilst code reviews from inexperienced developers tend towards more trivial issues such as naming conventions and visual representation.
Although inexperienced reviewers do discuss functional issues, the majority of issues were obvious syntactic bugs e.g., "\textit{Does this even work with a `;` inside those brackets?}", whilst experienced reviewers largely focused on deeper issues that require explanations e.g., "\textit{Can we make sure these two calls (ciphers and applicationProtocolConfig), or at least applicationProtocolConfig call, are made after the customizer is applied? Otherwise we would end up violating HTTP/2 specification IIUC}".
In general, the findings from our preliminary analysis coincides with past research, which further motivates the ELF method.

\section{Results}
In this section, we discuss the results with respect to the three RQs, which evaluate accuracy (5.1), informativeness (5.2) and issue types discussed (5.3) in the comments generated by our experience-aware loss functions (ELF) method.

For the count based manual evaluation results on the 100 random samples, we are interested in measuring any significant improvements from our models compared to the original CodeReviewer model.
To this end, we elect to report statistical significance based on the one-tailed two proportion Z-test~\cite{fleiss2013}.
The two proportion Z-test is used for comparing two proportions for significant differences, in this case one proportion would be the result of the original CodeReviewer model and the other proportion would be the result of one of our models.
We select a one-tailed test as we are only interested in an improvement in one direction, e.g., significant increase in the number of suggestions generated out of 100 samples compared to the original CodeReviewer model.

\subsection{(RQ1) What is the impact of ELF on the accuracy of code review comment generation models?}
Below, we present the accuracy results of our ELF models based on BLEU-4, semantic equivalence, and applicability.

\textbf{\underline{BLEU-4.}}  \textit{\textbf{ All ELF models surpassed past methods in terms of matched \textbf{n}-grams.
The best-performing ELF models achieved 
5\% higher BLEU-4 scores than the original CodeReviewer model.}}
As shown in Table~\ref{tab:accuracy}, all ELF models (with different ownership values at different granularity levels) achieved higher BLEU-4 scores than the original CodeReviewer.
In contrast, the experience-aware oversampling models achieved lower BLEU-4 scores.
Comparing the performance of ELF models across different ownership values and granularities, all the models achieved comparable BLEU-4 scores ranging from 7.29 to 7.6.
The highest performing ELF models were $\omega_{aco\_Repo}$ and $\omega_{avg\_Repo}$, both recording without stop word BLEU-4 scores of 7.6, which is a $+5\%$ increase over the original
CodeReviewer model.
The results are also consistent when considering stop words in BLEU-4.
These results suggest that ELF models can generate more comments that are textually similar to the ground truth than past models.

\textbf{\underline{Semantic Equivalence.}} \textit{\textbf{ Our ELF models achieved results comparable to those of the original CodeReviewer model in terms of generating semantically equivalent comments to the ground truth.
}}
Table~\ref{tab:accuracy} shows that 20 comments out of 100 samples generated by the original CodeReviewer model are semantically equivalent to the ground truth.
Our ELF models also achieve similar results, ranging from
17 to 23 comments that are semantically equivalent to the ground truth
, where seven of 12 ELF models generated more than 20 of such comments.
In contrast, all experience-aware oversampling models generated less than 20 semantically equivalent comments.
The highest performing ELF models were $\omega_{rso\_Repo}$ and $\omega_{rso\_Pkg}$, both achieving 23 correct matches.
These results suggest that the use of experience-aware methods does not have a large impact on the semantic equivalence of model generated comments towards the ground truth.
However, it is important to note that this evaluation did not consider the quality of generated comments.
Given that our key goal is to improve the quality of generated comments by aligning with the comments of experienced reviewers, we did not expect to achieve an improvement in terms of semantic equivalence towards the general population of the dataset.
To this end, we assess the quality of generated comments using the subsequent metrics.

\textbf{\underline{Applicability.}} \textit{\textbf{ 
Our ELF models can generate more comments that are applicable to the code changes than the original CodeReviewer model. 
The top performing ELF model generated 29\% more applicable comments than the original CodeReviewer model.}}
Table~\ref{tab:accuracy} shows that 42 comments out of 100 samples generated by the original CodeReviewer model were applicable to the code change.
All ELF models apart from $\omega_{aco\_Repo}$ generated more applicable comments than the original CodeReviewer model, ranging from 43 to 54 applicable comments.
For experience-aware oversampling, only $MA$ could outperform the original CodeReviewer model.
In total, eight of 12 ELF models generated more applicable comments than all past techniques.
Comparing across ELF strategies, we found that $\omega_{rso}$ models were consistently high performing, generating between 52 and 53 applicable comments.
Overall, $\omega_{aco\_Sys}$ was the top performer with 54 applicable comments, which is a statistically significant increase of +29\% over the original CodeReviewer model. 

\begin{mdframed}[linewidth=0.5mm,roundcorner=12pt, backgroundcolor=black!10]
\textbf{RQ1:} All ELF models surpassed past methods in terms of BLEU-4, achieving up to +5\% increase over the original CodeReviewer model.
Overall, all ELF models achieved comparable results to the original CodeReviewer model in terms of semantically equivalent comments.
Our ELF models were able to generate more applicable comments than all past methods, achieving up to +29\% increase over the original CodeReviewer model.

\end{mdframed}

\begin{table}[]
\caption{BLEU-4 on the Entire Test Set, Semantic Equivalence \& Applicablility on 100 Random Samples}
\label{tab:accuracy}
\resizebox{\columnwidth}{!}{%
\begin{tabular}{|l|l|l|lll|lll|lll|lll|lll|}
\hline

 & &  & \multicolumn{3}{c|}{Exp-aware Oversampling} & \multicolumn{12}{c|}{Exp-aware  Loss Function (ELF)} \\ 
 
& &  & \multicolumn{3}{c|}{$ $}  & \multicolumn{3}{c}{$\omega_{aco}$} & \multicolumn{3}{c}{$\omega_{rso}$} & \multicolumn{3}{c}{$\omega_{avg}$} & \multicolumn{3}{c|}{$\omega_{max}$} \\ \hline
 
 & & \textit{ORG} & \textit{MRMA} & \textit{MR} & \textit{MA} & \textit{Repo} & \textit{Sys} & \textit{Pkg} & \textit{Repo} & \textit{Sys} & \textit{Pkg} & \textit{Repo} & \textit{Sys} & \textit{Pkg} & \textit{Repo} & \textit{Sys} & \textit{Pkg} \\ \hline

 \multirow{2}{*}{$Test\phantom{0}Set$} & $B4_{w/o \phantom{0} Stop words}$ & \underline{7.27} & 6.87 \textcolor{red}{$\downarrow$} & 6.71 \textcolor{red}{$\downarrow$} & 7.11 \textcolor{red}{$\downarrow$} & \textbf{7.6}\phantom{-} \textcolor{ForestGreen}{$\uparrow$} & 7.56 \textcolor{ForestGreen}{$\uparrow$} & 7.46 \textcolor{ForestGreen}{$\uparrow$} & \phantom{0}7.45 \textcolor{ForestGreen}{$\uparrow$} & 7.57 \textcolor{ForestGreen}{$\uparrow$} & 7.55  \textcolor{ForestGreen}{$\uparrow$} & \phantom{0}\textbf{7.6}\phantom{-} \textcolor{ForestGreen}{$\uparrow$} & 7.36 \textcolor{ForestGreen}{$\uparrow$} & 7.45 \textcolor{ForestGreen}{$\uparrow$} & 7.43 \textcolor{ForestGreen}{$\uparrow$} & 7.29 \textcolor{ForestGreen}{$\uparrow$} & 7.38 \textcolor{ForestGreen}{$\uparrow$} \\

  & $B4_{w/ \phantom{0} Stop words}$  & \underline{5.81} &  5.59 \textcolor{red}{$\downarrow$}  & 5.5\phantom{0} \textcolor{red}{$\downarrow$} & 5.72 \textcolor{red}{$\downarrow$} & 6.09 \textcolor{ForestGreen}{$\uparrow$} & 6.07 \textcolor{ForestGreen}{$\uparrow$} & 6.02 \textcolor{ForestGreen}{$\uparrow$} & \phantom{0}5.92 \textcolor{ForestGreen}{$\uparrow$} & \textbf{6.1}\phantom{-} \textcolor{ForestGreen}{$\uparrow$}  & 6.02 \textcolor{ForestGreen}{$\uparrow$} & \phantom{0}6.07 \textcolor{ForestGreen}{$\uparrow$}  & 5.95 \textcolor{ForestGreen}{$\uparrow$} & 5.96 \textcolor{ForestGreen}{$\uparrow$} & 5.93 \textcolor{ForestGreen}{$\uparrow$} & 5.82 \textcolor{ForestGreen}{$\uparrow$} & 5.99 \textcolor{ForestGreen}{$\uparrow$} \\
\hline

 
 \multirow{2}{*}{$Samples$} & SE & \phantom{.0}\underline{20} & \phantom{.0}14 \textcolor{red}{$\downarrow$} & \phantom{.0}18 \textcolor{red}{$\downarrow$} & \phantom{.0}19 \textcolor{red}{$\downarrow$} & \phantom{.0}17 \textcolor{red}{$\downarrow$} & \phantom{.0}21 \textcolor{ForestGreen}{$\uparrow$} & \phantom{.0}19 \textcolor{red}{$\downarrow$} & \phantom{--0}\textbf{23} \textcolor{ForestGreen}{$\uparrow$} & \phantom{.-}\textbf{23} \textcolor{ForestGreen}{$\uparrow$} & \phantom{.0}20 & \phantom{.00}22 \textcolor{ForestGreen}{$\uparrow$} &\phantom{.0}20 & \phantom{.0}22 \textcolor{ForestGreen}{$\uparrow$} & \phantom{.0}22 \textcolor{ForestGreen}{$\uparrow$} & \phantom{.0}22 \textcolor{ForestGreen}{$\uparrow$} & \phantom{.0}19 \textcolor{red}{$\downarrow$}  \\ 

 & App & \phantom{.0}\underline{42} & \phantom{.0}40 \textcolor{red}{$\downarrow$} & \phantom{.0}38 \textcolor{red}{$\downarrow$} & \phantom{.0}44 \textcolor{ForestGreen}{$\uparrow$} & \phantom{.0}37 \textcolor{red}{$\downarrow$} & \phantom{..}$\bf54^{*}$\textcolor{ForestGreen}{$\uparrow$} & \phantom{.0}53 \textcolor{ForestGreen}{$\uparrow$} & \phantom{.00}52 \textcolor{ForestGreen}{$\uparrow$} & \phantom{.0}53 \textcolor{ForestGreen}{$\uparrow$} & \phantom{.0}53 \textcolor{ForestGreen}{$\uparrow$} & \phantom{.00}44 \textcolor{ForestGreen}{$\uparrow$} & \phantom{.0}43 \textcolor{ForestGreen}{$\uparrow$} & \phantom{.0}46 \textcolor{ForestGreen}{$\uparrow$} & \phantom{.0}48 \textcolor{ForestGreen}{$\uparrow$} & \phantom{.0}46 \textcolor{ForestGreen}{$\uparrow$} & \phantom{.0}44 \textcolor{ForestGreen}{$\uparrow$} \\ 
\hline

\multicolumn{16}{l}{\footnotesize Original Code Reviewer (\textit{ORG}), Major Reviewer Major Author (\textit{MRMA}), Major Reviewer (\textit{MR}), Major Author (\textit{MA}), Repository (\textit{Repo}), Subsystem (\textit{Sys}), Package (\textit{Pkg})}  \\
\multicolumn{16}{l}{\footnotesize BLEU-4 (B4), Semantic Equivalence (SE), Applicablility (App)}  \\
\multicolumn{16}{l}{\footnotesize Increased from \textit{ORG} (\textcolor{ForestGreen}{$\uparrow$}), Decreased from \textit{ORG} (\textcolor{red}{$\downarrow$}), \textit{p}<0.05 ($^*$)} \\
\end{tabular}
}
\end{table}

\begin{table}[]
\caption{Feedback Type and Presence of Explanation on 100 Random Samples}
\label{tab:information}
\resizebox{\columnwidth}{!}{%
\begin{tabular}{|l|l|l|lll|lll|lll|lll|lll|}
\hline

 &  &  & \multicolumn{3}{c|}{\begin{tabular}[c]{@{}c@{}}Exp-aware \end{tabular}} & \multicolumn{12}{c|}{\begin{tabular}[c]{@{}c@{}}Exp-aware Loss Function (ELF)\end{tabular}} \\  
 &  &  &   \multicolumn{3}{c|}{\begin{tabular}[c]{@{}c@{}}Oversampling \end{tabular}}   & \multicolumn{3}{c}{$\omega_{aco}$} & \multicolumn{3}{c}{$\omega_{rso}$} & \multicolumn{3}{c}{$\omega_{avg}$} & \multicolumn{3}{c|}{$\omega_{max}$} \\ \hline
 
 & \textit{GT} & \textit{ORG} & \textit{MRMA} & \textit{MR} & \textit{MA} & \textit{Repo} & \textit{Sys} & \textit{Pkg} & \textit{Repo} & \textit{Sys} & \textit{Pkg} & \textit{Repo} & \textit{Sys} & \textit{Pkg} & \textit{Repo} & \textit{Sys} & \textit{Pkg} \\ \hline
 
Sug & 87 & \underline{27} & $31^{\phantom{*}}$\textcolor{ForestGreen}{$\uparrow$} & $28^{\phantom{*}}$\textcolor{ForestGreen}{$\uparrow$} & $30^{\phantom{*}}$\textcolor{ForestGreen}{$\uparrow$} & $25^{\phantom{*}}$\textcolor{red}{$\downarrow$} & $34^{\phantom{*}}$\textcolor{ForestGreen}{$\uparrow$} & $\bf42^{*}$\textcolor{ForestGreen}{$\uparrow$} & 34 \textcolor{ForestGreen}{$\uparrow$} & $35^{\phantom{*}}$\textcolor{ForestGreen}{$\uparrow$} & 37 \textcolor{ForestGreen}{$\uparrow$} & 29 \textcolor{ForestGreen}{$\uparrow$} & $24^{\phantom{*}}$\textcolor{red}{$\downarrow$} & 31 \textcolor{ForestGreen}{$\uparrow$} & 30 \textcolor{ForestGreen}{$\uparrow$} & 28 \textcolor{ForestGreen}{$\uparrow$} & $30^{\phantom{*}}$\textcolor{ForestGreen}{$\uparrow$} \\ 

Con & 10 & \phantom{0}\underline{8} & \phantom{0}8 & \phantom{0}$9^{\phantom{*}}$\textcolor{ForestGreen}{$\uparrow$} & $11^{\phantom{*}}$\textcolor{ForestGreen}{$\uparrow$} & $10^{\phantom{*}}$\textcolor{ForestGreen}{$\uparrow$} & $\bf17^{*}$\textcolor{ForestGreen}{$\uparrow$} & \phantom{0}$9^{\phantom{*}}$\textcolor{ForestGreen}{$\uparrow$} & 14 \textcolor{ForestGreen}{$\uparrow$} & $15^{\phantom{*}}$\textcolor{ForestGreen}{$\uparrow$} & 11 \textcolor{ForestGreen}{$\uparrow$} & 11 \textcolor{ForestGreen}{$\uparrow$} & $16^{*}$\textcolor{ForestGreen}{$\uparrow$} & 12 \textcolor{ForestGreen}{$\uparrow$} & 12 \textcolor{ForestGreen}{$\uparrow$} & 14 \textcolor{ForestGreen}{$\uparrow$} & $12^{\phantom{*}}$\textcolor{ForestGreen}{$\uparrow$} \\

CQ & \phantom{0}3 & \phantom{0}\underline{7} & \phantom{0}$\bf1^{*}$\textcolor{red}{$\downarrow$} & \phantom{0}$\bf1^{*}$\textcolor{red}{$\downarrow$} & \phantom{0}$3^{\phantom{*}}$\textcolor{red}{$\downarrow$} & \phantom{0}$2^{*}$\textcolor{red}{$\downarrow$} & \phantom{0}$3^{\phantom{*}}$\textcolor{red}{$\downarrow$} &  \phantom{0}$2^{*}$\textcolor{red}{$\downarrow$} & \phantom{0}4 \textcolor{red}{$\downarrow$} & \phantom{0}$3^{\phantom{*}}$\textcolor{red}{$\downarrow$} & \phantom{0}5 \textcolor{red}{$\downarrow$} & \phantom{0}4 \textcolor{red}{$\downarrow$} & \phantom{0}$3^{\phantom{*}}$\textcolor{red}{$\downarrow$} & \phantom{0}3 \textcolor{red}{$\downarrow$} & \phantom{0}6 \textcolor{red}{$\downarrow$} & \phantom{0}4 \textcolor{red}{$\downarrow$} & \phantom{0}$2^{*}$\textcolor{red}{$\downarrow$} \\ \hline

Exp & 68 & \phantom{0}\underline{8} & $16^*$\textcolor{ForestGreen}{$\uparrow$} & $16^*$\textcolor{ForestGreen}{$\uparrow$} & $\bf20^*$\textcolor{ForestGreen}{$\uparrow$} & $10^{\phantom{*}}$\textcolor{ForestGreen}{$\uparrow$} & $11^{\phantom{*}}$\textcolor{ForestGreen}{$\uparrow$} & $15^{\phantom{*}}$\textcolor{ForestGreen}{$\uparrow$} & 11 \textcolor{ForestGreen}{$\uparrow$} & $18^*$\textcolor{ForestGreen}{$\uparrow$} & 11 \textcolor{ForestGreen}{$\uparrow$} & 11 \textcolor{ForestGreen}{$\uparrow$} & $12^{\phantom{*}}$\textcolor{ForestGreen}{$\uparrow$}  & 13 \textcolor{ForestGreen}{$\uparrow$}  & 12 \textcolor{ForestGreen}{$\uparrow$}  & 11 \textcolor{ForestGreen}{$\uparrow$}  & $15^{\phantom{*}}$\textcolor{ForestGreen}{$\uparrow$}  \\ \hline

\multicolumn{17}{l}{\footnotesize Ground Truth (\textit{GT}), Original Code Reviewer (\textit{ORG}), Major Reviewer Major Author (\textit{MRMA}), Major Reviewer (\textit{MR}), Major Author (\textit{MA})}  \\
\multicolumn{17}{l}{\footnotesize Repository (\textit{Repo}), Subsystem (\textit{Sys}), Package (\textit{Pkg}), Suggestion (Sug), Concern (Con), Confused Question (CQ), Explanation (Exp)}  \\
\multicolumn{17}{l}{\footnotesize Increased from \textit{ORG} (\textcolor{ForestGreen}{$\uparrow$}), Decreased from \textit{ORG} (\textcolor{red}{$\downarrow$}), \textit{p}<0.05 ($^*$)} \\
\end{tabular}
}
\end{table}

\subsection{(RQ2) What is the impact of ELF on the informativeness of code review comment generation models?}

\textbf{\underline{Feedback Type.}} \textit{\textbf{ Our ELF models provided more suggestions than all past methods.
The top performing ELF model generated 56\% more suggestions than the original CodeReviewer model.}}
Table~\ref{tab:information} shows that 27 comments out of 100 samples generated by the original CodeReviewer model were suggestions (64\% of its applicable comments).
With the exception of $\omega_{aco\_Repo}$ and $\omega_{avg\_Sys}$, all ELF models provided more suggestions than the original CodeReviewer model, ranging from 28 to 42 suggestions.
The experience-aware oversampling models also outperformed the original CodeReviewer model in terms of number of generated suggestions, however, the improvements were not significant.
Overall, five of 12 ELF models surpassed all past techniques in terms of generated suggestions, all of which were $\omega_{aco}$ and $\omega_{rso}$ models.
Comparing across strategies, we found that $\omega_{rso}$ models showed the most consistent improvements, generating between 34 and 37 suggestions.
Comparing across granularities, we found that package level models tended to provide the most suggestions, as they were consistently the highest performer across all four strategies for this task.
The top performer, $\omega_{aco\_Pkg}$, generated 42 suggestions (79\% of its applicable comments), yielding a statistically significant increase of +56\% over the original CodeReviewer model.

\textit{\textbf{All ELF models generated more concerns and fewer confused questions than the original CodeReviewer model. 
The top performing ELF models generated 71\% less confused questions than the original CodeReviewer model.}}
Table~\ref{tab:information} shows that eight comments out of 100 samples generated by the original CodeReviewer model were concerns.
All ELF models generated more concerns than the original CodeReviewer model, ranging from nine to 17 concerns generated.
Similarly, both $MR$ and $MA$ also provided more concerns than the original CodeReviewer model, however, the differences were not significant.
Interestingly, subsystem level views tended to produce the most concerns across all four strategies for this task.
Table~\ref{tab:information} shows that seven comments out of 100 samples generated by the original CodeReviewer model were confused questions (17\% of its applicable comments).
All ELF models manifested less confusion than the original CodeReviewer model, ranging from two to six confused questions generated.
The top performing ELF models, $\omega_{aco\_Repo}$, $\omega_{aco\_Pkg}$ and $\omega_{max\_Pkg}$ generated only two confused questions each (5\% of their applicable comments), which were statistically significant decreases of -71\% against the original CodeReviewer model.
Similarly, all experience-aware oversampling models also exhibited less confusion, with $MRMA$ and $MR$ generating only one confused question each.

\textbf{\underline{Presence of Explanation.}} \textit{\textbf{ We found that all ELF models tended to provide rationales more frequently than the original CodeReviewer model. 
The top performing ELF model generated 125\% more comments with rationales than the original CodeReviewer model.}}
Table~\ref{tab:information} shows that eight comments out of 100 samples generated by the original CodeReviewer model contained a rationale (19\% of its applicable comments).
All ELF models surpassed the original CodeReviewer model in this regard, generating between 10 to 18 comments with rationales.
Amongst the ELF models, $\omega_{rso\_Sys}$ was the top performer, generating 18 of such comments (34\% of its applicable comments), demonstrating a statistically significant increase of +125\% over the original CodeReviewer model.
Overall, experience-aware oversampling models still provided the most explanations, with $MA$ generating up to 20 comments with rationales.

\begin{mdframed}[linewidth=0.5mm,roundcorner=12pt, backgroundcolor=black!10]
\textbf{RQ2:} Our ELF models were able to provide more suggestions than all past methods, achieving up to +56\% increase over the original CodeReviewer model.
Similar to experience-aware oversampling, all ELF models also generated more concerns with fewer confused questions, achieving up to -71\% decrease compared to the original CodeReviewer model in terms of confused questions generated.
All ELF models provided explanations more frequently, achieving up to +125\% increase over the original CodeReviewer model.
\end{mdframed}

\subsection{(RQ3) What issue types are discussed in the ELF models' code reviews?}

\textbf{\underline{Functional Issues.}} \textit{\textbf{ Our ELF models generated more comments related to functional issues than all past methods.
The top performing model identified +129\% more functional issues than the original CodeReviewer model.}}
Table~\ref{table:category} shows that seven comments out of 100 samples generated by the original CodeReviewer model were related to functional issues (17\% of its applicable comments).
In this regard, nine of 12 ELF models surpassed all past methods, generating between 10 to 16 functional issue related comments.
In terms of experience-aware oversampling methods, both $MRMA$ and $MR$ outperformed the original CodeReviewer model, however the improvements were not significant. 
Comparing across the granularities, we found that package level models tended to provide more comments related to functional issues, however there was no observable pattern in the specific category types that yielded the improvement.
Compared to the original CodeReviewer model, we found that $\omega_{aco\_Pkg}$ was able to elicit five new comments related to high priority issues, such as logical, validation, and functional defects.
The top performing model for this task was $\omega_{avg\_Pkg}$, which generated 16 comments related to functional issues (35\% of its applicable comments), demonstrating a statistically significant increase of +129\% over the original CodeReviewer model.
Most of this improvement can be attributed to its ability to identify new logical errors and resource issues i.e., incorrect initialisation, manipulation and release.

\textbf{\underline{Evolvability Issues.}} \textit{\textbf{ Our ELF models generated more comments related to evolvability issues than the original CodeReviewer model. 
Specifically, all ELF models unanimously identified more documentation related issues than the original CodeReviewer model.
}}
Table~\ref{table:category} shows that 24 comments out of 100 samples generated by the original CodeReviewer model were related to evolvability issues.
In comparison, six of 12 ELF models generated more evolvability related comments, ranging from 27 to 29 of such comments.
For experience-aware oversampling, both $MRMA$ and $MA$ also slightly outperformed the original CodeReviewer model.
Comparing across strategies, we found that $\omega_{rso}$ models yielded the most consistent improvements, achieving a +13\% increase over the original code reviewer model.
These improvements can be attributed to their ability to identify opportunities for improving documentation and code element renaming.
Interestingly, every ELF model identified more documentation related issues, where $\omega_{aco}$ and $\omega_{rso}$ models all achieved statistically significant improvements, reaching up to +600\% increase over the original CodeReviewer model.
All ELF models produced fewer comments related to trivial issues, i.e., visual representation, achieving up to -50\% decrease from the original CodeReviewer model.
Overall, $\omega_{aco\_Pkg}$ yielded the most improvement, with a +21\% increase over CodeReviewer, identifying more improvement opportunities for a wide array of evolvability issues (four of six categories).

\textbf{\underline{Discussion.}} \textit{\textbf{ All ELF models outperformed past methods in terms of discussion invoking comments.
The top performing ELF models generated 150\% more questions concerning implementation choices than the original CodeReviewer model.}}
Table~\ref{table:category} shows that three comments out of 100 samples generated by the original CodeReviewer model were discussions.
In contrast, all ELF models outperformed past methods, generating between six to 11 discussions.
In this aspect, experience-aware oversampling demonstrated negligible overall difference compared to the original CodeReviewer model.
The majority of improvements from the ELF models can be attributed to an increase in questions concerning implementation choices. In particular, $\omega_{aco\_Sys}$, $\omega_{rso\_Sys}$ and $\omega_{avg\_Sys}$ exhibited statistically significant increases of +150\% over CodeReviewer in terms of these types of questions.

\begin{mdframed}[linewidth=0.5mm,roundcorner=12pt, backgroundcolor=black!10]
\textbf{RQ3:} Our ELF models identified more functional issues than all past methods, achieving up to +129\% increase over the original CodeReviewer model.
Our ELF models found more evolvability issues.
Specifically, all ELF models identified more opportunities for improving documentation, achieving up to +600\% increase over the original CodeReviewer model.
All ELF models generated more questions concerning implementation choices than past methods, achieving up to +150\% increase over the original CodeReviewer model.
\end{mdframed}

\definecolor{palepink}{rgb}{0.98, 0.85, 0.87}
\definecolor{papayawhip}{rgb}{1.0, 0.94, 0.84}
\definecolor{pastelblue}{rgb}{0.68, 0.78, 0.81}

\begin{table}[]
\caption{Code Review Comment Categories on 100 Random Samples}
\label{table:category}
\resizebox{\columnwidth}{!}{%
\begin{tabular}{|l|l|l|lll|lll|lll|lll|lll|}
\hline

 &  &  & \multicolumn{3}{c|}{\begin{tabular}[c]{@{}c@{}}Exp-aware \end{tabular}} & \multicolumn{12}{c|}{\begin{tabular}[c]{@{}c@{}}Exp-aware Loss Function (ELF)\end{tabular}} \\  
 &  &  &   \multicolumn{3}{c|}{\begin{tabular}[c]{@{}c@{}}Oversampling \end{tabular}}   & \multicolumn{3}{c}{$\omega_{aco}$} & \multicolumn{3}{c}{$\omega_{rso}$} & \multicolumn{3}{c}{$\omega_{avg}$} & \multicolumn{3}{c|}{$\omega_{max}$} \\ \hline
 
 & \textit{GT} & \textit{ORG} & \textit{MRMA} & \textit{MR} & \textit{MA} & \textit{Repo} & \textit{Sys} & \textit{Pkg} & \textit{Repo} & \textit{Sys} & \textit{Pkg} & \textit{Repo} & \textit{Sys} & \textit{Pkg} & \textit{Repo} & \textit{Sys} & \textit{Pkg} \\ \hline

\Xhline{3\arrayrulewidth}
\rowcolor{palepink}
\textbf{Total Functional Issues} & 35 & \phantom{0}\underline{7} & \phantom{0}9\phantom{.} \textcolor{ForestGreen}{$\uparrow$} & \phantom{0}9\phantom{0} \textcolor{ForestGreen}{$\uparrow$} & \phantom{0}7 & 10\phantom{.} \textcolor{ForestGreen}{$\uparrow$} & 12\phantom{0}\textcolor{ForestGreen}{$\uparrow$} & 13$\phantom{^{*}}$\textcolor{ForestGreen}{$\uparrow$} & 10$\phantom{^{*}}$\textcolor{ForestGreen}{$\uparrow$} & 11\phantom{0}\textcolor{ForestGreen}{$\uparrow$} & 12\phantom{0}\textcolor{ForestGreen}{$\uparrow$} & \phantom{0}8$\phantom{^{*}}$\textcolor{ForestGreen}{$\uparrow$} & \phantom{0}9$\phantom{^{*}}$\textcolor{ForestGreen}{$\uparrow$} & $\bf16^{*}$\textcolor{ForestGreen}{$\uparrow$} & 11 \textcolor{ForestGreen}{$\uparrow$} & \phantom{0}9$\phantom{^{*}}$\textcolor{ForestGreen}{$\uparrow$} & 11 \textcolor{ForestGreen}{$\uparrow$} \\ \Xhline{3\arrayrulewidth}
 
\rowcolor{palepink}
Functional Defect & \phantom{0}6 & \phantom{0}\underline{2} & \phantom{0}2 & \phantom{0}1\phantom{0} \textcolor{red}{$\downarrow$} & \phantom{0}3\phantom{0} \textcolor{ForestGreen}{$\uparrow$}  & \phantom{0}1\phantom{.} \textcolor{red}{$\downarrow$} & \phantom{0}3\phantom{0}\textcolor{ForestGreen}{$\uparrow$} & \phantom{0}\textbf{4}$\phantom{^{*}}$\textcolor{ForestGreen}{$\uparrow$} &  \phantom{0}3\phantom{$^{*}$}\textcolor{ForestGreen}{$\uparrow$} & \phantom{0}2 & \phantom{0}2 & \phantom{0}2 & \phantom{0}1\phantom{$^{*}$}\textcolor{red}{$\downarrow$} & \phantom{0}2  & \phantom{0}2 & \phantom{0}1\phantom{$^{*}$}\textcolor{red}{$\downarrow$} & \phantom{0}2 \\ 

\rowcolor{palepink}
Validation & \phantom{0}6 & \phantom{0}\underline{1}& \phantom{0}2\phantom{.} \textcolor{ForestGreen}{$\uparrow$} & \phantom{0}\textbf{4}$\phantom{^{*}}$ \textcolor{ForestGreen}{$\uparrow$} & \phantom{0}0\phantom{0} \textcolor{red}{$\downarrow$} & \phantom{0}1 & \phantom{0}1 & \phantom{0}2$\phantom{^{*}}$\textcolor{ForestGreen}{$\uparrow$} & \phantom{0}1 & \phantom{0}2$\phantom{^{*}}$\textcolor{ForestGreen}{$\uparrow$} & \phantom{0}1 & \phantom{0}0$\phantom{^{*}}$\textcolor{red}{$\downarrow$} & \phantom{0}2$\phantom{^{*}}$\textcolor{ForestGreen}{$\uparrow$} & \phantom{0}1 & \phantom{0}2 \textcolor{ForestGreen}{$\uparrow$} & \phantom{0}1 & \phantom{0}3 \textcolor{ForestGreen}{$\uparrow$} \\

\rowcolor{palepink}
Logical & \phantom{0}5 & \phantom{0}\underline{0} & \phantom{0}2\phantom{.} \textcolor{ForestGreen}{$\uparrow$} & \phantom{0}1\phantom{0} \textcolor{ForestGreen}{$\uparrow$} & \phantom{0}1\phantom{0} \textcolor{ForestGreen}{$\uparrow$} & \phantom{0}$3^{*}$\textcolor{ForestGreen}{$\uparrow$} & \phantom{0}$\bf4^{*}$\textcolor{ForestGreen}{$\uparrow$}  &  \phantom{0}2$\phantom{^{*}}$\textcolor{ForestGreen}{$\uparrow$} & \phantom{0}1\phantom{$^{*}$}\textcolor{ForestGreen}{$\uparrow$} & \phantom{0}2\phantom{$^{*}$}\textcolor{ForestGreen}{$\uparrow$} & \phantom{0}2\phantom{$^{*}$}\textcolor{ForestGreen}{$\uparrow$} & \phantom{0}2$\phantom{^{*}}$\textcolor{ForestGreen}{$\uparrow$} & \phantom{0}2$\phantom{^{*}}$\textcolor{ForestGreen}{$\uparrow$} &  \phantom{0}$3^{*}$\textcolor{ForestGreen}{$\uparrow$} &\phantom{0}2 \textcolor{ForestGreen}{$\uparrow$} & \phantom{0}2$\phantom{^{*}}$\textcolor{ForestGreen}{$\uparrow$} & \phantom{0}0 \textcolor{red}{$\downarrow$} \\ 

\rowcolor{palepink}
Interface & \phantom{0}7 & \phantom{0}\underline{1} & \phantom{0}0\phantom{.} \textcolor{red}{$\downarrow$} & \phantom{0}1 & \phantom{0}0\phantom{0} \textcolor{red}{$\downarrow$} & \phantom{0}0\phantom{.} \textcolor{red}{$\downarrow$}  &  \phantom{0}0\phantom{0}\textcolor{red}{$\downarrow$} & \phantom{0}\textbf{2}$\phantom{^{*}}$\textcolor{ForestGreen}{$\uparrow$} & \phantom{0}1 & \phantom{0}0$\phantom{^{*}}$\textcolor{red}{$\downarrow$} & \phantom{0}1 & \phantom{0}0$\phantom{^{*}}$\textcolor{red}{$\downarrow$} & \phantom{0}1 & \phantom{0}\textbf{2}\phantom{]}\textcolor{ForestGreen}{$\uparrow$} & \phantom{0}1 & \phantom{0}0$\phantom{^{*}}$\textcolor{red}{$\downarrow$} & \phantom{0}\textbf{2}\phantom{.}\textcolor{ForestGreen}{$\uparrow$}  \\ 

\rowcolor{palepink}
Resource & \phantom{0}6 & \phantom{0}\underline{2} & \phantom{0}2 & \phantom{0}0\phantom{0} \textcolor{red}{$\downarrow$}  & \phantom{0}2 & \phantom{0}3\phantom{.} \textcolor{ForestGreen}{$\uparrow$} & \phantom{0}2\phantom{0} & \phantom{0}1$\phantom{^{*}}$\textcolor{red}{$\downarrow$} & \phantom{0}3$\phantom{^{*}}$\textcolor{ForestGreen}{$\uparrow$} & \phantom{0}3$\phantom{^{*}}$\textcolor{ForestGreen}{$\uparrow$} & \phantom{0}5$\phantom{^{*}}$\textcolor{ForestGreen}{$\uparrow$}& \phantom{0}3$\phantom{^{*}}$\textcolor{ForestGreen}{$\uparrow$} & \phantom{0}2 & \phantom{0}\textbf{6}\phantom{]}\textcolor{ForestGreen}{$\uparrow$} & \phantom{0}3 \textcolor{ForestGreen}{$\uparrow$} & \phantom{0}4$\phantom{^{*}}$\textcolor{ForestGreen}{$\uparrow$} & \phantom{0}2 \\ 

\rowcolor{palepink}
Support & \phantom{0}1 & \phantom{0}\underline{\bf1} & \phantom{0}\bf1 & \phantom{0}\bf1 & \phantom{0}\bf1 & \phantom{0}\bf1 & \phantom{0}\bf1 & \phantom{0}\bf1 & \phantom{0}\bf1 & \phantom{0}\bf1 & \phantom{0}\bf1 & \phantom{0}\bf1 & \phantom{0}\bf1 & \phantom{0}\bf1 & \phantom{0}\bf1 & \phantom{0}\bf1 & \phantom{0}\bf1 \\ 

\rowcolor{palepink}
Timing & \phantom{0}4 & \phantom{0}\underline{0} & \phantom{0}0 & \phantom{0}\bf1$\phantom{^{*}}$ \textcolor{ForestGreen}{$\uparrow$} & \phantom{0}0 & \phantom{0}\bf1\phantom{]}\textcolor{ForestGreen}{$\uparrow$} & \phantom{0}\bf1\phantom{]}\textcolor{ForestGreen}{$\uparrow$} & \phantom{0}\bf1$\phantom{^{*}}$\textcolor{ForestGreen}{$\uparrow$}& \phantom{0}0 & \phantom{0}\bf1\phantom{]}\textcolor{ForestGreen}{$\uparrow$} & \phantom{0}0 & \phantom{0}0 & \phantom{0}0 &  \phantom{0}\bf1\phantom{]}\textcolor{ForestGreen}{$\uparrow$} & \phantom{0}0 & \phantom{0}0 & \phantom{0}\bf1\phantom{.}\textcolor{ForestGreen}{$\uparrow$} \\

\Xhline{3\arrayrulewidth}
\rowcolor{papayawhip}
\textbf{Total Evolvability Issues} & 57 & \underline{24} & 26\phantom{.} \textcolor{ForestGreen}{$\uparrow$} & 23\phantom{0} \textcolor{red}{$\downarrow$} & 28\phantom{0} \textcolor{ForestGreen}{$\uparrow$} & 19$\phantom{^{*}}$\textcolor{red}{$\downarrow$} & 27\phantom{0}\textcolor{ForestGreen}{$\uparrow$} & \textbf{29}$\phantom{]}$\textcolor{ForestGreen}{$\uparrow$}  & 27$\phantom{^{*}}$\textcolor{ForestGreen}{$\uparrow$}  & 27$\phantom{^{*}}$\textcolor{ForestGreen}{$\uparrow$} & 27$\phantom{^{*}}$\textcolor{ForestGreen}{$\uparrow$} & 24 & 21$\phantom{^{*}}$\textcolor{red}{$\downarrow$} & 19$\phantom{^{*}}$\textcolor{red}{$\downarrow$}& 23 \textcolor{red}{$\downarrow$} & 27$\phantom{^{*}}$\textcolor{ForestGreen}{$\uparrow$} & 23 \textcolor{red}{$\downarrow$} \\ \Xhline{3\arrayrulewidth}

\rowcolor{papayawhip}
Solution Approach & \phantom{0}7 & \phantom{0}\underline{4} & \phantom{0}2\phantom{.} \textcolor{red}{$\downarrow$} & \phantom{0}3\phantom{0} \textcolor{red}{$\downarrow$} & \phantom{0}4 & \phantom{0}1$\phantom{^{*}}$\textcolor{red}{$\downarrow$} &  \phantom{0}3\phantom{0}\textcolor{red}{$\downarrow$} & \phantom{0}2\phantom{$^{*}$}\textcolor{red}{$\downarrow$}  & \phantom{0}3$\phantom{^{*}}$\textcolor{red}{$\downarrow$} & \phantom{0}4 & \phantom{0}3\phantom{$^{*}$}\textcolor{red}{$\downarrow$} &  \phantom{0}3$\phantom{^{*}}$\textcolor{red}{$\downarrow$} & \phantom{0}3$\phantom{^{*}}$\textcolor{red}{$\downarrow$} & \phantom{0}2\phantom{$^{*}$}\textcolor{red}{$\downarrow$} & \phantom{0}3 \textcolor{red}{$\downarrow$}  & \phantom{0}3$\phantom{^{*}}$\textcolor{red}{$\downarrow$} & \phantom{0}\textbf{5}\phantom{.}\textcolor{ForestGreen}{$\uparrow$}  \\

\rowcolor{papayawhip}
Documentation & \phantom{0}7 & \phantom{0}\underline{0} & \phantom{0}$3^{*}$\textcolor{ForestGreen}{$\uparrow$} & \phantom{0}2\phantom{0} \textcolor{ForestGreen}{$\uparrow$} & \phantom{0}2\phantom{0} \textcolor{ForestGreen}{$\uparrow$} & \phantom{0}$3^{*}$\textcolor{ForestGreen}{$\uparrow$} & \phantom{0}$4^{*}$\textcolor{ForestGreen}{$\uparrow$} & \phantom{0}$4^{*}$\textcolor{ForestGreen}{$\uparrow$} & \phantom{0}$4^{*}$\textcolor{ForestGreen}{$\uparrow$} & \phantom{0}$\bf6^{*}$\textcolor{ForestGreen}{$\uparrow$} & \phantom{0}$\bf6^{*}$\textcolor{ForestGreen}{$\uparrow$} &  \phantom{0}$3^{*}$\textcolor{ForestGreen}{$\uparrow$} & \phantom{0}2$\phantom{^{*}}$\textcolor{ForestGreen}{$\uparrow$} & \phantom{0}1$\phantom{^{*}}$\textcolor{ForestGreen}{$\uparrow$} & \phantom{0}2 \textcolor{ForestGreen}{$\uparrow$} & \phantom{0}$3^{*}$\textcolor{ForestGreen}{$\uparrow$}  & \phantom{0}2 \textcolor{ForestGreen}{$\uparrow$} \\ 

\rowcolor{papayawhip}
Organisation of Code & 20 & \phantom{0}\underline{3} & \phantom{0}6\phantom{.} \textcolor{ForestGreen}{$\uparrow$} & \phantom{0}3 &  \phantom{0}$\bf9^{*}$ \textcolor{ForestGreen}{$\uparrow$} & \phantom{0}2$\phantom{^{*}}$\textcolor{red}{$\downarrow$} & \phantom{0}2\phantom{0}\textcolor{red}{$\downarrow$} &  \phantom{0}5$\phantom{^{*}}$\textcolor{ForestGreen}{$\uparrow$} & \phantom{0}4$\phantom{^{*}}$\textcolor{ForestGreen}{$\uparrow$} & \phantom{0}3 & \phantom{0}2$\phantom{^{*}}$\textcolor{red}{$\downarrow$} & \phantom{0}5$\phantom{^{*}}$\textcolor{ForestGreen}{$\uparrow$}  & \phantom{0}5$\phantom{^{*}}$\textcolor{ForestGreen}{$\uparrow$}  & \phantom{0}2$\phantom{^{*}}$\textcolor{red}{$\downarrow$} & \phantom{0}4 \textcolor{ForestGreen}{$\uparrow$} &  \phantom{0}7$\phantom{^{*}}$\textcolor{ForestGreen}{$\uparrow$} & \phantom{0}2 \textcolor{red}{$\downarrow$} \\ 

\rowcolor{papayawhip}
Alternate Output & \phantom{0}6 & \phantom{0}\underline{4} & \phantom{0}3\phantom{.} \textcolor{red}{$\downarrow$} & \phantom{0}4 & \phantom{0}2\phantom{0} \textcolor{red}{$\downarrow$} & \phantom{0}1$\phantom{^{*}}$\textcolor{red}{$\downarrow$} & \phantom{0}\textbf{7}\phantom{0}\textcolor{ForestGreen}{$\uparrow$} &  \phantom{0}6$\phantom{^{*}}$\textcolor{ForestGreen}{$\uparrow$} & \phantom{0}3$\phantom{^{*}}$\textcolor{red}{$\downarrow$} & \phantom{0}2$\phantom{^{*}}$\textcolor{red}{$\downarrow$}& \phantom{0}4 & \phantom{0}1$\phantom{^{*}}$\textcolor{red}{$\downarrow$} & \phantom{0}2$\phantom{^{*}}$\textcolor{red}{$\downarrow$} & \phantom{0}4 & \phantom{0}4 & \phantom{0}1$\phantom{^{*}}$\textcolor{red}{$\downarrow$} & \phantom{0}2 \textcolor{red}{$\downarrow$} \\ 

\rowcolor{papayawhip}
Naming Convention & 12 & \phantom{0}\underline{7} & \phantom{0}\bf9\phantom{]}\textcolor{ForestGreen}{$\uparrow$} & \phantom{0}7\phantom{0} \textcolor{ForestGreen}{$\uparrow$} & \phantom{0}4\phantom{0} \textcolor{red}{$\downarrow$} & \phantom{0}7 &  \phantom{0}6\phantom{0}\textcolor{red}{$\downarrow$} & \phantom{0}8$\phantom{^{*}}$\textcolor{ForestGreen}{$\uparrow$} & \phantom{0}\bf9\phantom{-}\textcolor{ForestGreen}{$\uparrow$} & \phantom{0}\bf9$\phantom{^{*}}$\textcolor{ForestGreen}{$\uparrow$} & \phantom{0}\bf9$\phantom{^{*}}$\textcolor{ForestGreen}{$\uparrow$} & \phantom{0}7 & \phantom{0}5$\phantom{^{*}}$\textcolor{red}{$\downarrow$} & \phantom{0}7 & \phantom{0}7 & \phantom{0}\bf9\phantom{]}\textcolor{ForestGreen}{$\uparrow$} & \phantom{0}8 \textcolor{ForestGreen}{$\uparrow$} \\ 

\rowcolor{papayawhip}
Visual Representation & \phantom{0}5 & \phantom{0}\underline{6} & \phantom{0}3\phantom{.} \textcolor{red}{$\downarrow$} & \phantom{0}4\phantom{0} \textcolor{red}{$\downarrow$} & \phantom{0}\bf7$\phantom{^{*}}$ \textcolor{ForestGreen}{$\uparrow$} & \phantom{0}5$\phantom{^{*}}$\textcolor{red}{$\downarrow$} & \phantom{0}5\phantom{0}\textcolor{red}{$\downarrow$} & \phantom{0}4$\phantom{^{*}}$\textcolor{red}{$\downarrow$} & \phantom{0}4$\phantom{^{*}}$\textcolor{red}{$\downarrow$} & \phantom{0}3$\phantom{^{*}}$\textcolor{red}{$\downarrow$} & \phantom{0}3$\phantom{^{*}}$\textcolor{red}{$\downarrow$} & \phantom{0}5$\phantom{^{*}}$\textcolor{red}{$\downarrow$}  & \phantom{0}4$\phantom{^{*}}$\textcolor{red}{$\downarrow$} & \phantom{0}3$\phantom{^{*}}$\textcolor{red}{$\downarrow$} & \phantom{0}3 \textcolor{red}{$\downarrow$} & \phantom{0}4$\phantom{^{*}}$\textcolor{red}{$\downarrow$} & \phantom{0}4 \textcolor{red}{$\downarrow$} \\

\Xhline{3\arrayrulewidth}
\rowcolor{pastelblue}
\textbf{Total Discussion}& \phantom{0}3 & \phantom{0}\underline{4} & \phantom{0}4 & \phantom{0}5\phantom{0} \textcolor{ForestGreen}{$\uparrow$} & \phantom{0}5\phantom{0} \textcolor{ForestGreen}{$\uparrow$} & \phantom{0}6$\phantom{^{*}}$\textcolor{ForestGreen}{$\uparrow$} & $\bf11^{*}$\textcolor{ForestGreen}{$\uparrow$} & \phantom{0}8$\phantom{^{*}}$\textcolor{ForestGreen}{$\uparrow$} & $10^{*}$\textcolor{ForestGreen}{$\uparrow$} & $\bf11^{*}$\textcolor{ForestGreen}{$\uparrow$} & \phantom{0}8$\phantom{^{*}}$\textcolor{ForestGreen}{$\uparrow$} & \phantom{0}7$\phantom{^{*}}$\textcolor{ForestGreen}{$\uparrow$} & $10^{*}$\textcolor{ForestGreen}{$\uparrow$} & \phantom{0}7$\phantom{^{*}}$\textcolor{ForestGreen}{$\uparrow$} & \phantom{0}7 \textcolor{ForestGreen}{$\uparrow$} & \phantom{0}6$\phantom{^{*}}$\textcolor{ForestGreen}{$\uparrow$} & \phantom{0}8 \textcolor{ForestGreen}{$\uparrow$} \\ \Xhline{3\arrayrulewidth}

\rowcolor{pastelblue}
Question & \phantom{0}1 & \phantom{0}\underline{4} & \phantom{0}4 & \phantom{0}2\phantom{0} \textcolor{red}{$\downarrow$} & \phantom{0}4\phantom{0} & \phantom{0}6$\phantom{^{*}}$\textcolor{ForestGreen}{$\uparrow$} & $\bf10^{*}$\textcolor{ForestGreen}{$\uparrow$} & \phantom{0}7$\phantom{^{*}}$\textcolor{ForestGreen}{$\uparrow$} & \phantom{0}9$\phantom{^{*}}$\textcolor{ForestGreen}{$\uparrow$} & $\bf10^{*}$\textcolor{ForestGreen}{$\uparrow$} & \phantom{0}7$\phantom{^{*}}$\textcolor{ForestGreen}{$\uparrow$} & \phantom{0}6$\phantom{^{*}}$\textcolor{ForestGreen}{$\uparrow$} & $\bf10^{*}$\textcolor{ForestGreen}{$\uparrow$} & \phantom{0}7$\phantom{^{*}}$\textcolor{ForestGreen}{$\uparrow$} & \phantom{0}7 \textcolor{ForestGreen}{$\uparrow$} & \phantom{0}6$\phantom{^{*}}$\textcolor{ForestGreen}{$\uparrow$} & \phantom{0}6 \textcolor{ForestGreen}{$\uparrow$} \\ 

\rowcolor{pastelblue}
Design Discussion & \phantom{0}2 & \phantom{0}\underline{0} & \phantom{0}0 & \phantom{0}$\bf3^{*}$ \textcolor{ForestGreen}{$\uparrow$} & \phantom{0}1\phantom{0} \textcolor{ForestGreen}{$\uparrow$} & \phantom{0}0 & \phantom{0}1\phantom{0}\textcolor{ForestGreen}{$\uparrow$} & \phantom{0}1$\phantom{^{*}}$\textcolor{ForestGreen}{$\uparrow$} & \phantom{0}1$\phantom{^{*}}$\textcolor{ForestGreen}{$\uparrow$} & \phantom{0}1$\phantom{^{*}}$\textcolor{ForestGreen}{$\uparrow$} & \phantom{0}1$\phantom{^{*}}$\textcolor{ForestGreen}{$\uparrow$} & \phantom{0}1$\phantom{^{*}}$\textcolor{ForestGreen}{$\uparrow$} & \phantom{0}0 & \phantom{0}0 & \phantom{0}0 & \phantom{0}0 & \phantom{0}2 \textcolor{ForestGreen}{$\uparrow$} \\ \Xhline{3\arrayrulewidth}

\rowcolor{lightgray}
\textbf{Other} & \phantom{0}2 & \phantom{0}\underline{0} &  \phantom{0}0 & \phantom{0}0 & \phantom{0}\bf1$\phantom{^{*}}$ \textcolor{ForestGreen}{$\uparrow$} & \phantom{0}0 & \phantom{0}\bf1\phantom{]}\textcolor{ForestGreen}{$\uparrow$}  & \phantom{0}\bf1$\phantom{^{*}}$\textcolor{ForestGreen}{$\uparrow$}  & \phantom{0}\bf1$\phantom{^{*}}$\textcolor{ForestGreen}{$\uparrow$} & \phantom{0}\bf1$\phantom{^{*}}$\textcolor{ForestGreen}{$\uparrow$} & \phantom{0}\bf1$\phantom{^{*}}$\textcolor{ForestGreen}{$\uparrow$} & \phantom{0}\bf1$\phantom{^{*}}$\textcolor{ForestGreen}{$\uparrow$} & \phantom{0}0 & \phantom{0}\bf1$\phantom{^{*}}$\textcolor{ForestGreen}{$\uparrow$} & \phantom{0}\bf1 \textcolor{ForestGreen}{$\uparrow$} & \phantom{0}0 & \phantom{0}0 \\ \hline

\multicolumn{17}{l}{\footnotesize Ground Truth (\textit{GT}), Original Code Reviewer (\textit{ORG}), Major Reviewer Major Author (\textit{MRMA}), Major Reviewer (\textit{MR}), Major Author (\textit{MA})}  \\
\multicolumn{17}{l}{\footnotesize Repository (\textit{Repo}), Subsystem (\textit{Sys}), Package (\textit{Pkg}), Functional Issue \fcolorbox{black}{palepink}{\null} Evolvability Issue \fcolorbox{black}{papayawhip}{\null} Discussion \fcolorbox{black}{pastelblue}{\null} }  \\
\multicolumn{17}{l}{\footnotesize Increased from \textit{ORG} (\textcolor{ForestGreen}{$\uparrow$}), Decreased from \textit{ORG} (\textcolor{red}{$\downarrow$}), \textit{p}<0.05 ($^{*}$)} \\
\end{tabular}
}
\end{table}

\section{Analysis and Discussion}

In this section, we provide further analysis of our results.
Firstly, we investigate the distribution of ownership values and how they are related to each other (7.1).
Secondly, we compare the quality of code reviews generated by ELF against real code reviews (7.2).
Thirdly, we discuss the BLEU-4 score improvements of our ELF models (7.3).
Next, we discuss the value of considering reviewer experience at different granularities (7.4).
Finally, we compare the performances of different ELF strategies (7.5) and conduct a sensitivity analysis (7.6).

\begin{figure}
    \centering
    \includegraphics[width=0.8\textwidth]{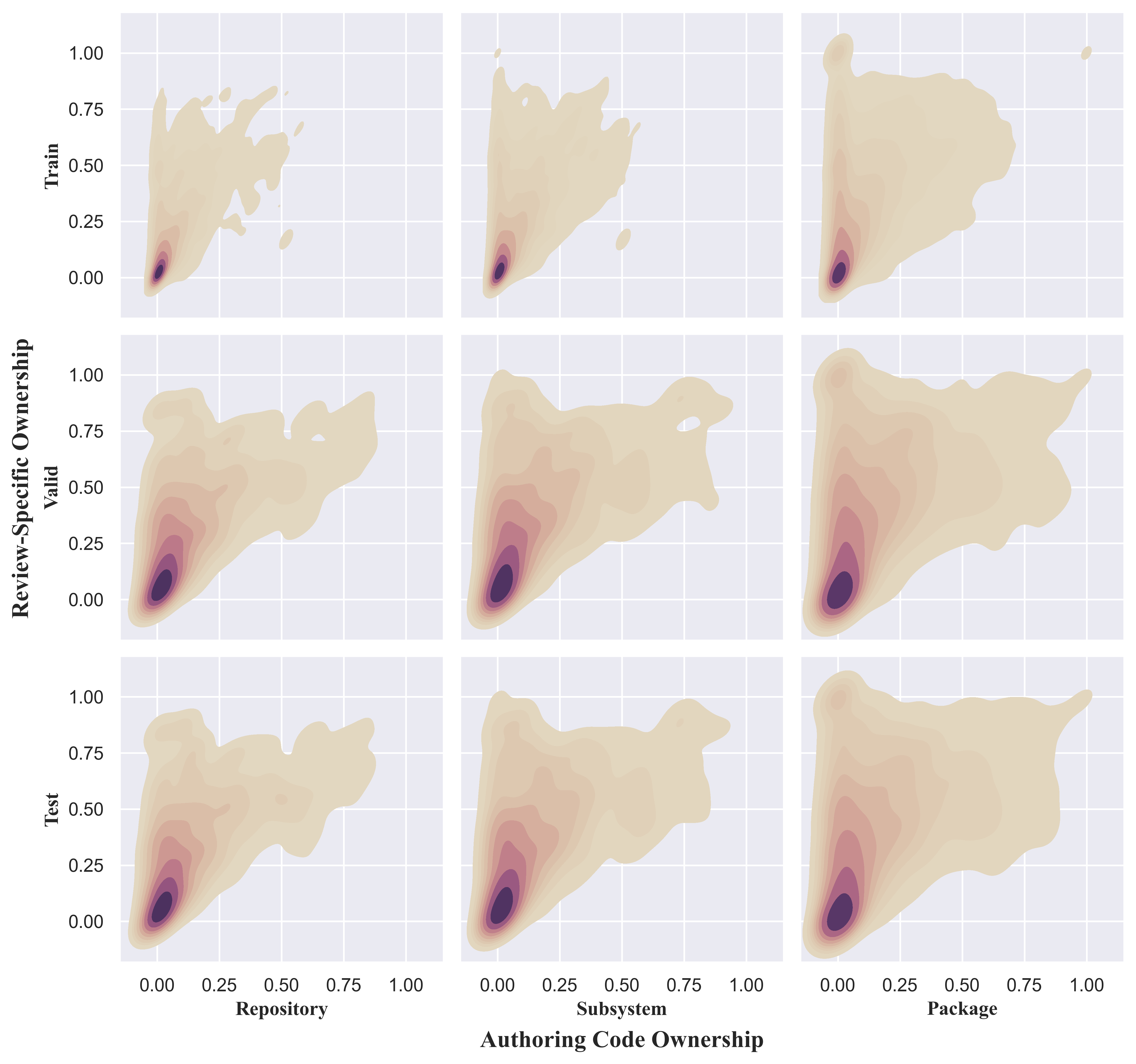}
    \caption{Kernel Density Estimates of ACO and RSO at Repository, Subsystem and Package Level}
    \label{fig:kde}
    \Description[kde]{KDE plots}
\end{figure}

\subsection{How are the different ownership values distributed and what is their relationship to each other?}
A difference in behaviour amongst the ELF models is only possible if reviewers' ownership ratios vary between the authoring and reviewing perspective, and across the different granularities.
Thus, we first investigate the distributions and relationships of the different ownership ratios to gain initial insight into why different ELF models may generate different code review comments.

The kernel density estimates of the ownership ratios in Figure~\ref{fig:kde} demonstrate that reviewers tend to have higher RSO than ACO.
In fact, many reviews are provided by experienced developers who contribute mainly via reviewing code, rather than by writing code, which aligns with past findings~\cite{rso}.
Interestingly, a rise in ACO values consistently comes with a rise in RSO values, indicating that developers who are responsible for a large portion of commits tend to also be responsible for a larger portion of code reviews.
Comparing across the dataset splits, it is evident that ownership ratios in the training set are more concentrated around smaller values as opposed to the validation and test sets. 
This aligns with intuition since it is more difficult for an individual to gain high ownership coverage in large projects that have more contributors.
We find that ACO and RSO values consistently increase as we consider a more refined granularity, indicating that many developers have more specialised coverage within the project.
We report the Pearson correlation for ACO in the training set between repository and subsystem ($\rho$=0.85), between subsystem and package ($\rho$=0.69) and finally between repository and package ($\rho$=0.58).
We also report the Pearson correlation for RSO in the training set between repository and subsystem ($\rho$=0.67), between subsystem and package ($\rho$=0.74) and finally between repository and package ($\rho$=0.67).
\faLightbulbO \phantom{ }The divergence in both correlation and magnitude of ownership ratios between the different granular views indicate potential for varying signals.
As such, a difference in model behaviour when employing ELF from different granularities was within our expectation.

\subsection{How do code reviews generated by ELF compare with the ground truth in terms of quality?}

Whilst our main results section compares reviews generated by the different review comment generation models, this section compares the reviews generated by the best performing models to the ground truth reviews provided by human reviewers.
The 100 ground truth examples were randomly sampled from the distribution of the test set, thus it includes reviewers from a variety of experience levels.
This is particularly relevant as code reviewing in reality is not restricted to experienced reviewers only.
The main reason for this design was to compare the nature of code reviews generated from the different techniques against the nature of real code reviews that developers are receiving.
We report the summary statistics for the ownership values of these 100 samples (RSO Repository $(\mu, \sigma)$ = (0.26, 0.24), ACO Repository $(\mu, \sigma)$ = (0.14, 0.23), RSO Subsystem $(\mu, \sigma)$ = (0.31, 0.26), ACO Subsystem $(\mu, \sigma)$ = (0.16, 0.24), RSO Package $(\mu, \sigma)$ = (0.35, 0.29), ACO Package $(\mu, \sigma)$ = (0.19, 0.26)).
In terms of feedback type, the proportions of suggestions in the 100 samples is 87\%.
For the best performing ELF strategy in this regard, $\omega_{aco\_Pkg}$, only 79\% of their applicable comments were suggestions.
The proportion of code reviews that had explanations in the 100 samples is 68\%.
For the best performing ELF strategy in this regard, $\omega_{rso\_Sys}$, only 34\% of their applicable comments contained explanations.
In terms of comment categories, 35\% of the 100 samples discuss critical functional issues.
For the best performing ELF strategy in this regard, $\omega_{avg\_Pkg}$, 35\% of their applicable comments also discuss critical functional issues.
Although the ELF method yields similar behaviour with the real code reviews in terms of proportion of functional issues discussed, they still fall short in terms of informativeness.
Thus, reaching this informativeness standard still remains an objective for current models.
Future work should also progressively track model progress against more experienced reviewers.

\subsection{Why do ELF models generate more textually similar reviews to the ground truth?}
To investigate the unanimous improvement of ELF models in terms of BLEU-4, we manually analysed the top 10 examples for each ELF model in terms of the largest BLEU-4 delta against the original CodeReviewer model.

\begin{figure}
    \centering
    \includegraphics[width=1\textwidth]{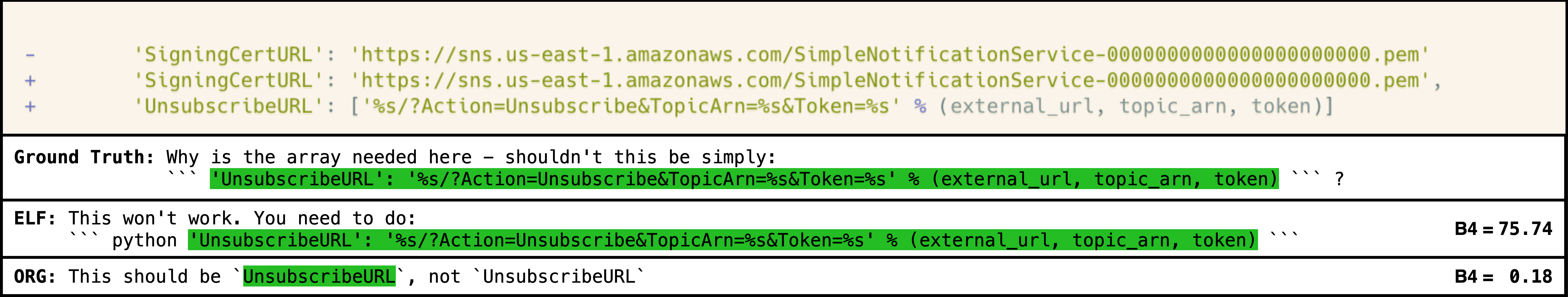}
    \caption{Example of an ELF Model Generating a Natural Language Code Review with Embedded Code Snippet}
    \label{fig:CS1}
    \Description[CS]{Code Snippet}
\end{figure}

\faLightbulbO \phantom{ }We observe that all ELF models tended to provide code snippets with exact implementation suggestions embedded within their natural language comment as opposed to the original model which tended to provide natural language comments that occasionally included code elements.
Thus, when ELF models were semantically equivalent to the ground truth, they often provided close to exact match code implementations to explain their suggestions, which achieves near perfect BLEU-4 results, as demonstrated in Figure~\ref{fig:CS1}.
In the cases where ELF models did not achieve semantic equivalence, they were able to locate the exact code that was considered problematic by the ground truth comment, which also resulted in high \textit{n}-gram matches.
This finding is significant given that using code snippets to explain implementations is rarely done in code reviews and a method used by experienced reviewers to guide new developers in the project~\cite{zhang2024demystifying}.
Often used for suggestions and citations, reviewers employ these code snippets when the intended revision cannot be clearly described in words. 
This ensures that ideas can be quickly evaluated with less room for miscommunication.
Given that these types of comments are well received by developers, we consider this an improvement in the quality of generated comments.
We emphasise that the ELF models have learned this behaviour without losing the ability to generate natural language reviews; this can be contrasted with the previously discussed issue where models learned only to copy and paste code inputs due to problematic examples in the dataset that included only code suggestions without any natural language component.

\subsection{What is the value of considering reviewer experiences from different granularities?}

As observed in the results, all granularities perform similarly in terms of BLEU-4, semantic equivalence and applicability.
However, similar metric results do not indicate that the different views are generating the same comments.
To further explore the value of employing different granularities, we examine whether there is a difference in generated comments within each of the four proposed strategies.

\begin{figure}
    \centering
    \includegraphics[width=0.9\textwidth]{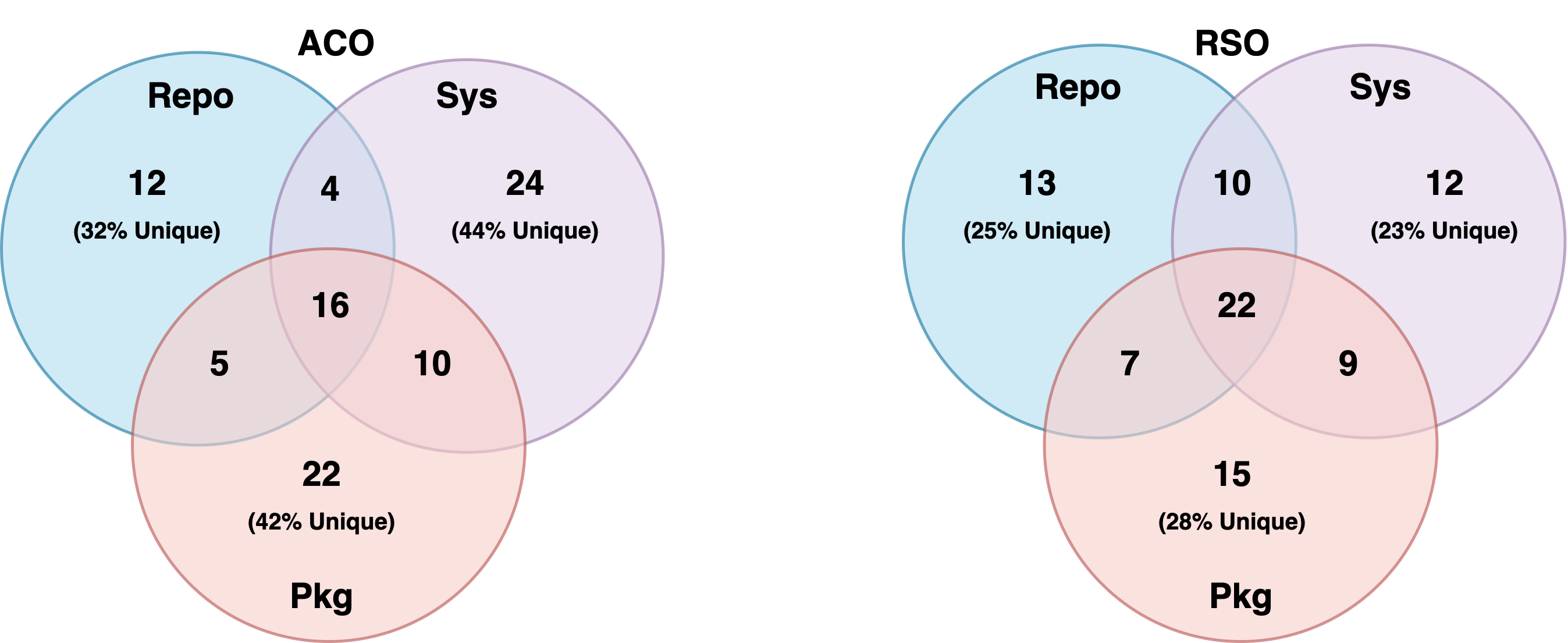}
    \caption{Diversity of Applicable Comments Generated by ACO and RSO Based Strategies in Terms of Comment Category}
    \label{fig:cov}
    \Description[cov]{coverage}
\end{figure}

Firstly, we investigate the degree to which different granularities are achieving semantic equivalence on the same ground truth examples.
For $\omega_{aco}$, we find that the three granular views can cover 29 semantically equivalent comments together with $\frac{10}{29}$ being mutually inclusive and $\frac{11}{29}$ attributable to only one unique view. 
For $\omega_{rso}$, we find that the three granular views can cover 34 semantically equivalent comments together with $\frac{10}{34}$ being mutually inclusive and $\frac{12}{34}$ attributable to only one unique view. 
For $\omega_{avg}$, we find that the three granular views can cover 32 semantically equivalent comments together with $\frac{10}{32}$ being mutually inclusive and $\frac{10}{32}$ attributable to only one unique view. 
For $\omega_{max}$, we find that the three granular views can cover 32 semantically equivalent comments together with $\frac{12}{32}$ being mutually inclusive and $\frac{13}{32}$ attributable to only one unique view. 
\faLightbulbO \phantom{ }For all strategies, more than half of the total semantically equivalent matches cannot be simultaneously covered by all three granular views, a substantial portion of which can only be covered by one particular view.
To further explore the diversity of the generated comments, we conducted the same analysis on the types of elicited comments, which is displayed in Figure~\ref{fig:cov}.
Two applicable code reviews of the same code change submission are considered to be different if they fall under different code review categories.
For $\omega_{aco}$, we find that 32\%, 44\% and 42\% of the applicable comments generated by repository, subsystem and package level views, respectively, were completely unique.
For $\omega_{rso}$, the proportion of applicable comments that were completely unique are 25\%, 23\% and 28\%, respectively.
\faLightbulbO \phantom{ }The high percentage of uniquely generated comments indicates that the varying ownership ratios between the granularities in fact do elicit diverging perspectives.
As such, we find that there is value in considering all three granularities when training automated code review models with ELF.

\subsection{Which ELF strategies elicit the most improvement in terms of code review comment quality?}
Comparing across the strategies, we find that all three models of $\omega_{rso}$ are consistently high performing across the majority of the tasks, i.e., semantic equivalence, applicability, suggestions provided, identified evolvability issues and
discussion invoking comments.
However, the subsystem and package level models of $\omega_{aco}$ are consistently high performing across nearly all tasks apart from semantic equivalence.
\faLightbulbO \phantom{ }In contrast, we find that both $\omega_{avg}$ and $\omega_{max}$ strategies rarely bring additional value, indicating that experience types should be considered separately during training.

Since the subsystem and package level models of both $\omega_{aco}$ and $\omega_{rso}$ are top performing, we analyse their overlap in issue types discussed in the generated comments.
We find that the overlap between $\omega_{aco\_Pkg}$ and $\omega_{rso\_Pkg}$ is only 22\% compared to the overlap between $\omega_{aco\_Sys}$ and $\omega_{rso\_Sys}$ at 37\%.
\faLightbulbO \phantom{ }This indicates that the package level models of $\omega_{aco}$ and $\omega_{rso}$ are not only more accurate and informative, but offer the most diverse range of code reviews.
Compared to CodeReviewer, $\omega_{aco\_Pkg}$ is able to identify more high priority issues related to functional defects, validation errors, and logical faults, which are the top three most useful code review categories as rated by open-source developers~\cite{turzo2024}.
We highlight the significance of this finding as code reviews rarely find functionality defects in reality~\cite{dontfindbugs}.
This demonstrates that targeting reviewers' coding experience at the package level can pinpoint critical code reviews during model training.
To complement $\omega_{aco\_Pkg}$, $\omega_{rso\_Pkg}$ is able to catch more resource-related issues.
$\omega_{aco\_Pkg}$ improves on a wide variety of evolvability issues, i.e. documentation, organisation of code (refactoring), alternate output and naming convention.
Unlike visual representation issues (formatting), these evolvability issues are often beyond the scope of traditional static analysis tools~\cite{vijayvergiya2024ai}, making them highly valuable types of code reviews to generate.
More specifically, improving the organisation of code can help ameliorate low evolvability in systems, which hinders developer productivity when adding features or fixing bugs~\cite{rombach1987,bandi2003}.
On the other hand, $\omega_{rso\_Pkg}$ excels at improving documentation, which can aid the comprehensibility of the programs~\cite{oman1990}.
Overall, we find that both $\omega_{aco}$ and $\omega_{rso}$ strategies are the most effective, especially at the package level.
Given the diverse nature of these two models, it is highly synergetic to integrate both models together.

\subsection{How sensitive is the ELF method to weighting strategy selection?}

\begin{figure}
    \centering
    \includegraphics[width=1\textwidth]{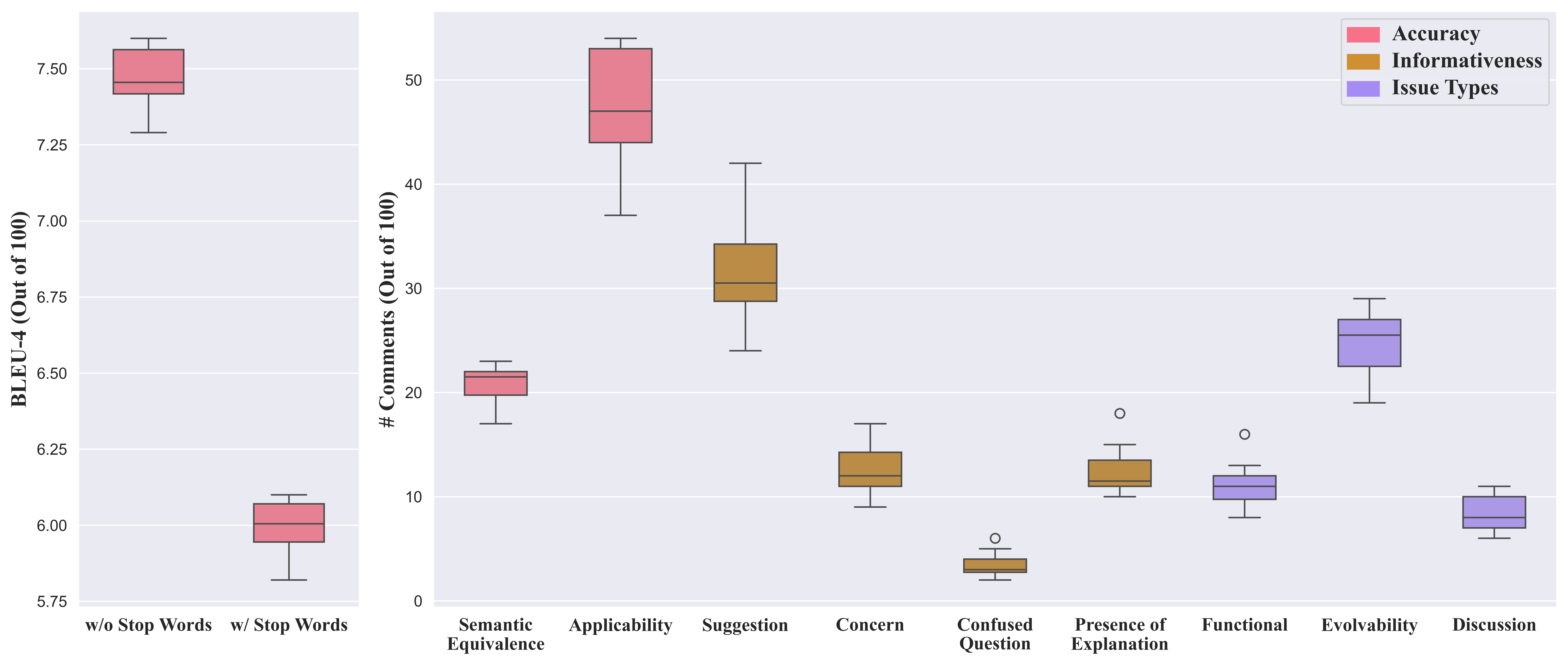}
    \caption{Distribution of Results Across All ELF Strategies}
    \label{fig:sensitivity}
    \Description[sen]{sensitivity}
\end{figure}

Although the ELF method removes the need to tune ownership thresholds, there still exists the need to select a particular loss function weighting.
Through our results, we identified $\omega_{aco\_Pkg}$ and $\omega_{rso\_Pkg}$ to be top performers, however, it remains unclear how sensitive the ELF method is to strategy selection overall.
To this end, we analyse the performance gaps between the different weighting strategies, using the same statistical testing as the main results.
Figure~\ref{fig:sensitivity} shows the distribution of results across all ELF weighting strategies.
When not considering the aforementioned top performers, we find that results can be sensitive to weighting strategy selection in terms of informativeness and issues types discussed, whilst accuracy is generally consistent.
We discuss findings from the sensitivity analysis in more detail below.

To reflect the impact of weighting strategy selection on accuracy, we report the gap between the best and worst performing ELF models in terms of BLEU-4, semantic equivalence and applicability.
The distribution of accuracy results are presented through the pink boxplots in Figure~\ref{fig:sensitivity}.
In terms of BLEU-4, we find that the differences in performance are minimal, where the maximum gap in BLEU-4 with and without stop words are merely $\Delta0.28$ (6.1 - 5.82) and $\Delta0.31$ (7.6 - 7.29), respectively.
Similarly, we find statistically insignificant differences in semantic equivalence, where the maximum gap is $\Delta6$ (23 - 17) in correct matches.
In contrast, the $\omega_{aco\_Repo}$ variant yields a statistically significant decrease of $\Delta17$ (54 - 37) in applicable comments from the top performer, whilst the rest of the variants yield insignificant differences.
With the exception of the case mentioned above, we find that accuracy of the ELF method vary minimally to weighting strategy selection.

To reflect the impact of weighting strategy selection on informativeness, we report the gap between the best and worst performing ELF models in terms of feedback type and presence of explanation.
The distribution of informativeness results are presented through the brown boxplots in Figure~\ref{fig:sensitivity}.
In terms of feedback type, we find that the variation in both suggestions and concerns provided can be statistically significant, with a maximum gap of $\Delta18$ (42 - 24) in suggestions and $\Delta8$ (17 - 9) in concerns.
In contrast, we find that amount of confused questions generated only varied by a maximum of $\Delta4$ (6 - 2), which is insignificant.
Similarly, we find that the variation in amount of generated explanations is also statistically insignificant, with a maximum gap of $\Delta8$ (18 - 10).
Thus, the ELF method varies minimally to weighting strategy selection for certain informativeness measures, such as number of confused questions and explanations generated, however, this selection causes large variations in the number of suggestions and concerns generated, when not considering $\omega_{aco\_Pkg}$ and $\omega_{rso\_Pkg}$.

To reflect the impact of weighting strategy selection on issue types discussed, we report the gap between the best and worst performing ELF models in terms of the categories of generated comments.
These distributions are presented through the purple boxplots in Figure~\ref{fig:sensitivity}.
In terms of functional issues, we find that $\omega_{avg\_Repo}$ can yield a statistically significant decrease of $\Delta8$ (16 - 8) from the top performer, whilst the rest of the variants yield insignificant differences.
For evolvability issues, we find that variations can be significant, where the maximum gap is $\Delta10$ (29 - 19).
In contrast, the amount of generated discussions vary minimally, where the maximum gap is $\Delta5$ (11 - 6).
In general, the ELF method varies minimally to weighting strategy selection in terms of functional issues and discussions.
On the other hand, this selection can cause significant variations in number of comments generated regarding evolvability issues, when not considering $\omega_{aco\_Pkg}$ and $\omega_{rso\_Pkg}$.

Since behaviour in terms of various informativeness measures and issues types discussed can vary due to the selected weighting strategy, future researchers and practitioners should preliminary evaluate the different strategies on any validation set of interest before selection.
Nevertheless, $\omega_{aco\_Pkg}$ and $\omega_{rso\_Pkg}$ remain consistently top performing across all evaluation metrics in our experiments.

\section{Broader Implications}
We now discuss the broader implications of our study.

\textbf{Practitioners.}
In theory, if code review comment generation models could successfully mimic experienced reviewers in disseminating expert knowledge during the review, the usefulness of the tool would better align with preferences of developers~\cite{kononenko2016code}.
However, given the non-trivial nature of the task, the state of open source code review comment generation models is still far from practical, as exhibited by the large portion of non-applicable comments being generated.
Whilst our released model checkpoints can be directly deployed for inference with frameworks that support the T5 language model architecture, software developers and project maintainers should proceed with caution, given the tool is in early development and has not demonstrated completely reliable levels of accuracy.
Nevertheless, the ELF method presents a data efficient way to improve code review comment generation models, which is especially critical given the sparsity of such corpora.
For model developers who want to leverage ELF to train their code review comment generation models, there are two base requirements that need to be fulfilled, 1) abundant code review data from mature software projects, and 2) accurate version control history.
An abundance of code review data is needed such that generalisable patterns can be learned, and mature software projects are required such that experienced reviewers can exist.
For a single project, the amount of code review data from experienced reviewers is often insufficient, thus practitioners should incorporate mature projects that have similar characteristics to their target project e.g., application domain, software engineering standards, programming languages.
Whilst there is no established threshold for determining when a project can be considered mature, the rule of thumb used to collect the training set was more than 2.5k pull requests~\cite{codereviewer}.
In terms of accurate version control history, practitioners can benefit from projects where a complete record of the history is available, and the entirety of each reviewer's involvement in the project can be resolved to one profile. 
Additionally, a more accurate calculation of the subsystem and package level expertise can be achieved if the model developers use projects where the ground truth software architecture is known.
To implement ELF in a real workflow, model developers need to 1) extract all code reviews along with pull-request and commit histories from the version control systems of target projects, 2) pre-calculate the ownership ratios for each example, 3) train their code review comment generation model with the ELF methodology, 4) deploy the static model in their code review platform, and 5) repeat steps 1-4 with newer code reviews when performance degradation is observed.

Whilst open source code review comment generation models are still in the early stages of development, our findings highlight the importance of attending to examples involving expert feedback during model training.
As such, practitioners should consider integrating reviewer experience into future model development as it can help optimise the usefulness of such tools for the developers they assist.

\textbf{Researchers.}
Whilst our preliminary results show promising signs for integrating reviewer experience into the training process of code review comment generation models, there are various areas for improvement.
Firstly, the accuracy of subsystem and package level experience estimation is dependent on the ability to approximate ground truth software architectures.
However, manual recovery of these architectures is unfeasible given the number of projects needed for building state-of-the-art models, therefore our approach relied on a heuristic based approximation that can be suboptimal.
Future researchers should explore the potential integration of methods in automatic software architecture recovery, such as those that rely on data mining~\cite{montes98}, information retrieval~\cite{PoshyvanykGM12} or clustering~\cite{Wiggerts97,AnquetilL99b,MitchellM06}.
This can facilitate more accurate experience estimation for fine-grained ownership ratios, thus improving the reliability of the ELF method.
Secondly, our ability to holistically evaluate the techniques on the entire test set is restricted by the absence of reliable automatic evaluation methods.
As a result, our study relied on costly manual evaluation, which only scales to small samples. 
Future researchers should investigate reliable automatic methods to evaluate natural language code reviews in terms of semantic equivalence, applicability, informativeness measures and comment categories.
This can facilitate rapid development e.g., search optimal parameter configurations or loss function designs.
Additionally, the domain knowledge of experienced reviewers can often be project specific and dependent on contextual information beyond the input code hunk.
As a result, the limited contextual information presents a restrictive upper bound on the effectiveness of the ELF method. 
Future researchers should investigate ways to retrieve and incorporate relevant project information discussed by the code review comment.

Whilst this study integrates the notion of reviewer experience into the fine-tuning process of code review comment generation models, the idea can be applied to any AI for software engineering task involving human created artifacts.
The only condition is that developer characteristics must induce a meaningful difference in the artifact of concern.
Additionally, the use of ownership ratios is not restricted to weighting loss functions, and can be incorporated into any technique that uses quantifiable measures e.g., weighting the ranking in retrieval systems~\cite{rag} or weighted data sampling~\cite{chang2017active, needell2014}.
In summary, our study sheds light on the utility in considering the characteristics of developers behind the targeted artifacts, which is a perspective that should be considered more generally by all researchers working on AI for software engineering.

\section{Threats to Validity}
We now discuss the threats to the validity of our study.

\textbf{Internal Validity.}
To ensure that our representation of CodeReviewer is faithful to the original paper~\cite{codereviewer}, we utilised their exact replication package, pre-trained checkpoint, hyper-parameters, and training setup.
The only varying factors are the filtered dataset used for fine-tuning and the experience-aware techniques that we introduce.
Ownership values may be underestimated when records of reviews or commits are lost, users are unsearchable, users use multiple accounts, and when projects are rebased or deleted.
As BLEU-4 is not an adequate measure for comprehensively evaluating the correctness of the automated code review models, we not only employ manual evaluation to capture semantic equivalence with the ground truth, but we also assess the applicability of the comments irrespective of the ground truth.
To capture the informativeness of generated reviews, we conduct manual evaluation in terms of both feedback type and presence of explanation.
To capture the change in behaviour in terms of the subject of generated comments, we conduct manual evaluation on the review category type.
Since manual evaluation is prone to subjectivity and bias, the two annotators conducted the manual evaluations independently with the sources of the generated reviews masked.
All conflicts were resolved together until a satisfactory inter-rater agreement level was reached, resulting in a refined guideline that was used to complete the rest of the annotations.
Two additional reviewers then independently checked the final results for consistency.
To manage the scale of manual annotation, we sampled at a 95\% confidence level with a 10\% margin of error, rather than a more precise 5\% margin of error. 
Nevertheless, we reported statistically significant results under the one-tailed two proportion Z-test.
All research outputs are included in the replication package for transparency~\footnote{\url{https://zenodo.org/records/15400484}}.

\textbf{External Validity.}
Our experiments on model training focused on 519 of the most starred projects on GitHub, with more than 2,500 pull requests.
Thus, the behaviours elicited from our ELF technique may not generalise to smaller scale software projects or to developers in other code review environments, e.g., Gerrit or closed source development.
Additionally, the dataset consists of only inline code review comments, which dictates the nature of the discussion.
As such, these findings may not generalise to other types of code review comments, e.g., commit level or pull request level.

\section{Conclusion}
The field of code review comment generation has demonstrated the potential of deep learning based language models in automating the cognitively loaded task of code reviewing.
Whilst past studies have focused on technical improvements using techniques derived from the field of machine learning, our study explores the potential of leveraging the software engineering concept of reviewer experience to elicit higher quality code reviews from automated code review models.
Our proposed experience-aware loss function (ELF) method re-weights the training data using traditional ownership metrics that reflect a reviewer's authoring and reviewing experiences at the repository, subsystem, and package level.
Through both quantitative and qualitative evaluation, our results show that certain ELF models can surpass all past methods in terms of both accuracy (RQ1) and informativeness (RQ2), whilst also identifying more functional and evolvability issues (RQ3) in the process.
In particular, we found that considering both authoring and reviewing experiences separately at the package level was the most beneficial; however, the uniqueness of the comments elicited from both the repository and subsystem level view also demonstrate that different granularities are highly complementary.
We hope that our findings can inspire future work in review comment generation to also consider integrating established software engineering concepts and theories into the design of automated code review models.

\section*{Acknowledgment}
This research was supported by The University of Melbourne’s Research Computing Services and the Petascale Campus Initiative.
Patanamon Thongtanunam was supported by the Australian Research Council's Discovery Early Career Researcher Award (DECRA) funding scheme (DE210101091).

\bibliographystyle{ACM-Reference-Format}
\bibliography{references}

\end{document}